\documentclass[preprint2]{aastex}
\usepackage{rotating}
\usepackage{graphicx}
\usepackage{grffile}

\shortauthors{Crepp et al. 2011}

\begin{document}
\setcounter{secnumdepth}{3}
\title{The Dynamical Mass and Three-Dimensional Orbit of HR7672B: A Benchmark Brown Dwarf with High Eccentricity}
\author{Justin R. Crepp\altaffilmark{1}, John Asher Johnson\altaffilmark{1,2}, Debra A. Fischer\altaffilmark{3}, Andrew W. Howard\altaffilmark{4}, Geoffrey W. Marcy\altaffilmark{4}, Jason T. Wright\altaffilmark{5,6}, Howard Isaacson\altaffilmark{4}, Tabetha Boyajian\altaffilmark{7,8}, Kaspar von Braun\altaffilmark{2}, Lynne A. Hillenbrand\altaffilmark{1}, Sasha Hinkley\altaffilmark{1,9}, John M. Carpenter\altaffilmark{1}, John M. Brewer\altaffilmark{3}}
\altaffiltext{1}{Department of Astrophysics, California Institute of Technology, 1200 E. California Blvd., Pasadena, CA 91125} 
\altaffiltext{2}{NASA Exoplanet Science Institute, California Institute of Technology, MC 100-22, Pasadena, CA 91125}
\altaffiltext{3}{Department of Physics, Yale University, New Haven, CT 06511} 
\altaffiltext{4}{Department of Astronomy, University of California, Berkeley, CA 94720} 
\altaffiltext{5}{Department of Astronomy \& Astrophysics, The Pennsylvania State University, University Park, 16802}
\altaffiltext{6}{Center for Exoplanets and Habitable Worlds, The Pennsylvania State University, University Park, 16802} 
\altaffiltext{7}{Center for High Angular Resolution Astronomy, Department of Physics and Astronomy, Georgia State University, Atlanta, GA 30302}
\altaffiltext{8}{Hubble Fellow}
\altaffiltext{9}{Sagan Fellow}
\email{jcrepp@astro.caltech.edu} 

\begin{abstract}  
The companion to the G0V star HR7672 directly imaged by \citet{liu_02} has moved measurably along its orbit since the discovery epoch, making it possible to determine its dynamical properties. Originally targeted with adaptive optics because it showed a long-term radial velocity acceleration (trend), we have monitored this star with precise Doppler measurements and have now established a 24 year time baseline. The radial velocity variations show significant curvature (change in the acceleration) including an inflection point. We have also obtained a recent image of HR7672B with NIRC2 at Keck. The astrometry also shows curvature. In this paper, we use jointly-fitted Doppler and astrometric models to calculate the three-dimensional orbit and dynamical mass of the companion. The mass of the host star is determined using a direct radius measurement from CHARA interferometry in combination with high resolution spectroscopic modeling. We find that HR7672B has a highly eccentric, $e=0.50^{+0.01}_{-0.01}$, near edge-on, $i=97.3^{+0.4}_{-0.5}$ deg, orbit with semimajor axis, $a=18.3^{+0.4}_{-0.5}$ AU. The mass of the companion is $m=68.7^{+2.4}_{-3.1}M_J$ at the 68.2\% confidence level. HR7672B thus resides near the substellar boundary, just below the hydrogen-fusing limit. These measurements of the companion mass are independent of its brightness and spectrum and establish HR7672B as a rare and precious ``benchmark" brown dwarf with a well-determined mass, age, and metallicity essential for testing theoretical evolutionary models and synthetic spectral models. It is presently the only directly imaged L,T,Y-dwarf known to produce an RV trend around a solar-type star.
\end{abstract}

\section{INTRODUCTION}\label{sec:intro}
Brown dwarfs have complicated atmospheres. Unlike stars, for which it is possible to infer bulk physical properties from spectra alone, the emergent radiation from substellar objects is currently not well understood (e.g., \citealt{cushing_08}). More than a decade following the direct detection of the first L and T dwarfs \citep{becklin_zuckerman_88,nakajima_95,oppenheimer_95,rebolo_95,basri_96,kirkpatrick_99_lt}, and now faced with the recent detection of the first Y dwarfs \citep{cushing_11,liu_11,luhman_11}, theoretical spectral models are still undergoing major developments as they presently do not capture all of the relevant physics involved in shaping the spectra of cold bodies \citep{allard_97,marley_02,baraffe_03,burrows_06,saumon_08}. 

Factors that complicate the interpretation of brown dwarf spectra include the necessity to simultaneously model: the opacity of molecular species having millions of absorption lines, which results in an ill-defined continuum; the formation and settling of dust grains; non-equilibrium chemistry resulting from convective mixing; and temporal changes from weather patterns, among other phenomena \citep{cushing_08,marley_10}. Furthermore, the basic model input parameters, such as mass, radius, age, metallicity, and effective temperature, are often degenerate with one another, particularly for the faintest objects for which only broadband photometry or low resolution spectroscopy are available \citep{dupuy_09,janson_11,galicher_11}. In order to improve our understanding of low temperature atmospheres and guide the development of more sophisticated models, it is necessary to measure one or more of these physical properties independently of spectra and photometry. 

Substellar companions found orbiting nearby stars serve as useful laboratories for calibrating theoretical models. Properties of the host star are more readily measured and may be used to infer properties of the companion, such as chemical composition or possibly age, under the assumption that the star and brown dwarf formed from the same material at the same time \citep{pinfield_06}. A number of these ``benchmark" systems have recently been discovered \citep{lane_01,close_05,liu_07, ireland_08_BD,dupuy_09,bowler_09,biller_10_pztel,wahhaj_11}. 

It is also possible to measure the \emph{mass} of brown dwarfs (and extrasolar planets) using orbital dynamics. This has been accomplished in the past using the transit technique in combination with Doppler radial velocity (RV) observations \citep{stassun_06,johnson_11_lhs6343}. However, substellar companions with large semimajor axes and proximate distances may be imaged directly (spatially resolved from their host) using adaptive optics (AO) and high-contrast imaging technology \citep{marois_06,oppenheimer_hinkley_09,absil_mawet_10,crepp_11}. The companion sky-projected orbit can then be traced by measuring its position relative to the star over multiple epochs. This type of astrometry has been used for decades to characterize the orbits of binary stars \citep{aitken_19} and is significantly easier to realize in practice compared to monitoring the ``wobble" of a star in the sky relative to an inertial reference grid with $<0.1$ mas accuracy.

If Doppler measurements of the primary star are also obtained over a comparable time baseline, one can construct a three-dimensional orbit. Since the astrometry breaks the $sin(i)$ inclination degeneracy resulting from RV measurements alone, a Keplerian model that self-consistently combines both data sets provides an estimate of the companion true mass (e.g., \citealt{boden_06}). Knowing the mass of non-hydrogen-fusing objects is crucial because it governs their luminosity evolution in time \citep{stevenson_91,burrows_97}. 


It is commonly thought that the long orbital periods of wide-separation companions prohibits the calculation of dynamical masses. However, those with semimajor axes as large as $a\approx30$ AU orbiting nearby ($d\lesssim100$ pc) stars may be characterized in several years to tens of years, because the size of the astrometric uncertainties are $\sim$1-2 orders of magnitude smaller than the size of the orbit on the sky (thanks in part to the recent miniaturization in the physical size of detector pixels). Orbital solutions converge when both the astrometry and RVs become ``unique", which generally corresponds to showing curvature or a change in the acceleration (also known as the ``jerk").\footnote{See \citet{lu_09} for analysis of an equivalent problem involving stars orbiting the supermassive black hole in the center of the Galaxy. The orbits are well-characterized even though the periods are of order thousands of years. Likewise, the orbits of comets residing in the outer solar system may be determined with a time baseline of only several days \citep{bernstein_00}.} With accuracies of $\approx10$ mas (e.g., the size of a pixel in the NIRC2 camera at Keck -- PI: Keith Matthews), it is possible to calculate the inclination, as well as other parameters, in only several tenths of an orbital phase wrap (tens of degrees change in the true anomaly) depending on the orbit orientation and observing dates \citep{konopacky_10,currie_11_betapic,dupuy_11}. The constraints become tighter with time as the companion executes its motion along the ellipse. 

A natural method to identify promising high-contrast imaging targets is to select stars that show long-term Doppler accelerations (trends). This approach is convenient because RV programs have now established time baselines exceeding a decade. For example, the L$4.5\pm1.5$ dwarf companion to HR7672 ($=$Gl 779, $=$HD190406) was discovered by \citet{liu_02} using this technique. This G0V star was originally chosen for AO observations because it showed unambiguous evidence for the existence of a distant companion with substellar minimum mass: an $\approx \;$-24 m/s/yr acceleration over a 14 year baseline. HR7672B is currently the only directly imaged L,T,Y-dwarf known to produce an RV trend around a solar-type star.  

In this paper, we present an updated RV time series that shows significant orbital motion of the companion since the epoch of the direct imaging discovery. We have also obtained a new astrometric measurement with NIRC2 at Keck. Combining our observations with astrometry from the literature, and performing a joint-fit analysis to the full data set, we construct a three-dimensional orbit and calculate the dynamical mass of HR7672B with a fractional error of only 4\% using Newtonian dynamics (95\% confidence level). By explicitly connecting the mass of a brown dwarf to its radiative properties, the results may be used to inform both theoretical evolutionary models and synthetic spectral models. This study is afforded by an equally detailed characterization of HR7672A, for which we have modeled its high resolution spectrum to determine its surface gravity and metallicity, acquired a spectral energy distribution (SED) to determine its luminosity, and measured its radius directly using interferometry. 
 
\section{OBSERVATIONS}\label{sec:observations}

\subsection{Doppler Measurements}\label{sec:doppler}
HR7672 was initially observed with the 0.6m Coude Auxiliary Telescope (CAT) at Lick Observatory on September 9, 1987 UT. Several years of observations using the CAT as well as the Lick 3m (Hamilton) -- both of which feed the Coude echelle spectrograph -- revealed the star to have a linear RV trend \citep{cumming_99}. We also began Doppler monitoring of this star at Keck on June 2, 1997 UT using the HIgh-Resolution Echelle Spectrometer (HIRES; \citealt{vogt_94}). For each instrument, we used the iodine cell referencing method to measure precise RVs \citep{marcy_butler_92,butler_96}. Both data sets showed a consistent long-term acceleration. When \citet{liu_02} targeted this star with AO, finding that indeed a low-mass companion with a wide orbit was responsible for the trend, there was very little, if any, perceptible change in the acceleration. 

We have continued to monitor HR7672 at both Lick and Keck Observatories and have now established a 24 year time baseline. Our most recent RV data reveal significant curvature, making it possible to place tight constraints on the companion orbit. When combined with astrometry, this information is sufficient to converge to a unique orbital solution and companion mass, despite having coverage over a fraction of a single orbit cycle, because the signal-to-noise ratio in both data sets is high. 

\begin{figure*}[!t]
\begin{center}
\includegraphics[height=2.8in]{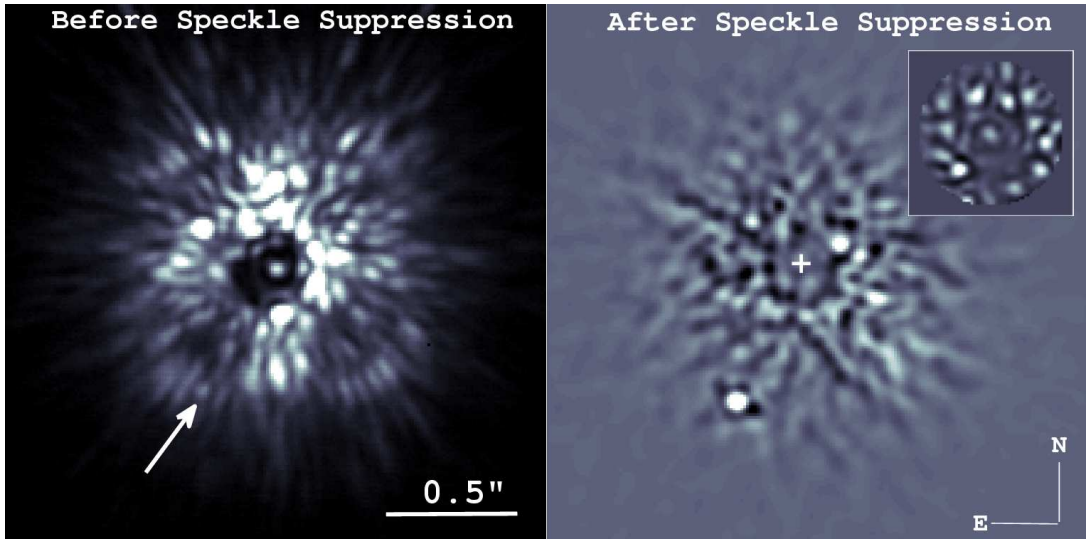} 
\caption{Direct image of HR7672B in the $K'$ filter taken on May 15, 2011 UT with NIRC2 at Keck. (left) Single image prior to PSF subtraction. The companion is located inside of the quasi-static speckle halo though discernible in individual frames. (right) Processed image using the ADI technique and LOCI algorithm. The companion is detected at 36$\sigma$ with 450 seconds of integration time. The inset subimage displays a cutout of the central region when our speckle suppression algorithm uses all available ADI frames but avoids PSF subtraction over the coronagraphic spot. The stellar Airy pattern is clearly visible facilitating astrometric measurements. We use data sets both with and without PSF subtraction to calculate the companion position relative to the star. This most recent observation increases the astrometric time baseline by 3.6 years.} 
\end{center}\label{fig:image}
\end{figure*} 

\subsection{Astrometry from High-Contrast Imaging}\label{sec:astrometry}
HR7672B has been imaged several times since its discovery. In July 2002, \citet{boccaletti_03} detected the companion with PHARO \citep{hayward_01} at Palomar in each of the $J$, $H$, $K_s$ bands to obtain color information and validate the spectral-type assigned by \citet{liu_02} from K-band spectroscopy. In September 2006, \citet{serabyn_09} observed HR7672B at Palomar to demonstrate the feasibility of new coronagraphic technologies and the use of ``extreme" AO. Both studies reported astrometric measurements of the position of HR7672B relative to the primary. We have also retrieved VLT archival images of the system from September 2007 (PI: A. Boccaletti -- program ID 279.C-5052(A)). These observations show clear orbital motion in a clockwise direction (north up, east left) when combined with the \citet{liu_02} results.  

We obtained additional images of HR7672B on May 15, 2011 using the Keck AO system with NIRC2. Given the large flux ratio between the primary and companion, we used the 300 mas diameter coronagraphic occulting spot to permit relatively long exposures without saturating the detector. The spot is partially transmissive and thus allows us to accurately measure the companion separation and position angle. Our initial data set consisted of 20 frames recorded using the narrow camera in position-angle mode (15 with the $K'$ filter and 5 with the $H$ filter). Although difficult to identify in a single exposure without prior knowledge of its approximate location, the majority of our raw frames did reveal the companion (Fig.~1). 

To further isolate the signal of HR7672 B from residual scattered starlight, we also obtained a 30 frame sequence with the $K'$ filter using the angular differential imaging (ADI) technique \citep{marois_06}. The parallactic angle changed by $12.7^{\circ}$ during this second sequence of images, providing sufficient angular diversity to remove the speckles and keep light from the companion \citep{lafreniere_loci}. Fig.~1 displays our high-contrast images before and after speckle suppression. 

\begin{table*}[!t]
\centerline{
\begin{tabular}{cccccc}
\hline
\hline
Year         &     JD-2,440,000      &    $\rho$ (mas)       &     PA ($^{\circ}$)         &   Instrument       &   Reference \\
\hline
2001.64   &  12144                &    $786 \pm 6$       &   $157.9\pm0.5$      &   NIRC2                  & \citet{liu_02}            \\
2001.94   &  12254                &    $794 \pm 5$       &   $157.3\pm0.6$      &   NIRC2                  & \citet{liu_02}            \\               
2002.54   &  12473                &    $788 \pm 6$       &   $156.6\pm0.9$      &   PHARO                & \citet{boccaletti_03}  \\  
2006.69   &  13989                &    $750 \pm 80$     &   $155.0\pm5.0$      &   PHARO               & \citet{serabyn_09}   \\
2007.73   &  14367                &    $742 \pm 35$     &   $151.8\pm2.9$      &   NACO                   & Program 279.C-5052(A)  \\
2011.37   &  15697                &    $519 \pm 6$       &   $147.1 \pm 0.5$    &   NIRC2                  &  present study          \\
\hline
\hline
\end{tabular}}
\caption{Astrometry measurements used for orbital analysis.}
\label{tab:astrom_meas}
\end{table*}

We measured the position of HR7672B relative to the primary star in each data set ($K'$ without ADI, $H$ without ADI, and $K'$ with ADI). The companion and stellar PSF core were fitted with a Gaussian function using least-squares iteration to find the astrometric centroid position in each processed frame. We correct for differential geometric distortion (warping) using the publicly available code provided by the Keck NIRC2 astrometry support webpage. Adopting a plate scale value of $9.963\pm0.006$ mas pix$^{-1}$ and instrument orientation relative to the sky of $0.13^{\circ}\pm0.02^{\circ}$ east of north, as measured by \citet{ghez_08}, we find a companion separation and position angle of $519\pm6$ mas and $147.1\pm0.5^{\circ}$ respectively.\footnote{This result is consistent with the values ($9.94 \pm 0.05$ mas pix$^{-1}$ and $0.0 \pm 0.5^{\circ}$) provided by the observatory. NIRC2 was not opened nor temperature cycled between 2003-2011, including the dates of our imaging observations (R. Campbell 2011, private communication).} HR7672B orbits in a clock-wise direction and is now significantly closer to its host star compared to the discovery epoch.

To estimate the uncertainty in our measurements, we calculate the standard deviation in the position resulting from each data set. We include a characteristic 5 mas error to account for astrometric bias introduced by the coronagraphic spot (Konopacky 2011, private communication). This error is added in quadrature with the uncertainty from the plate scale and instrument orientation. The results were combined using a weighted average to determine the final uncertainties. We find that the measurements from each data set are consistent with one another to within the ($1\sigma$) error bars. Simulated companions were injected into the images and their positions recovered to validate our results. Table~\ref{tab:astrom_meas} lists the separation and position angles used for our three-dimensional orbit analysis.\footnote{We do not include the original Gemini imaging data from June 2001, which suffers from systematic errors and a low signal-to-noise ratio as described in \citet{liu_02}.}

\subsection{Physical Properties of the Host Star}
The mass of the companion is tied to the mass of the primary star through the RV semi-amplitude and Kepler's equation. Astrometry determines the system total mass, and Doppler measurements can break the degeneracy between the system total mass and individual masses. However, we only have precise Doppler information for HR7672A, since the companion is significantly fainter than the primary. It is therefore important to reliably determine the mass of the host star independent of dynamical considerations. 

Fortunately, HR7672A is a nearby ($d=17.77\pm0.11$ pc) solar-type (G0V) star, making it possible to: (i) measure its radius directly using interferometry; and (ii) determine its surface gravity from high resolution spectra using models that are well-calibrated from observations of the Sun. We have spatially resolved the surface of HR7672A using the Georgia State University Center for High Angular Resolution astronomy Array (CHARA) \citep{ten05}. We have also obtained non-iodine (template) spectra of HR7672A for which to model. 

\begin{figure}[!t]								
  \begin{center}
      \includegraphics[height=2.35in]{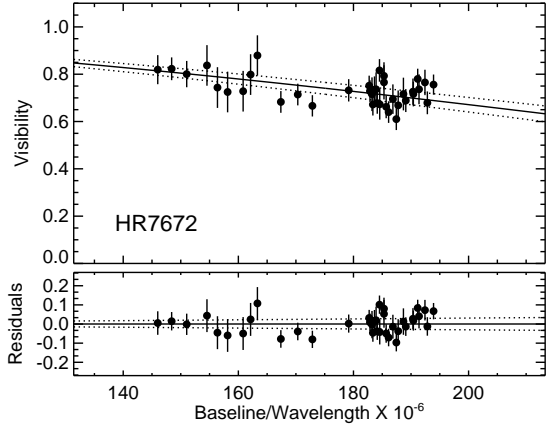}       
  \end{center}
  \caption{Calibrated visibility measurements and limb-darkened angular diameter fit for HR7672A. Dotted lines indicate the $3\sigma$ uncertainty level in the overall fit to the Bessell function. The bottom panel shows the residuals ($\chi^2_r=0.5$). Interferometric measurements of the radius of HR7672A allow us to place tight constraints on its mass.}
  \label{fig:visibility}
\end{figure}

In the following, we calculate the mass of HR7672A using an iterative version of the {\it Spectroscopy Made Easy} (SME) spectral modeling routine \citep{valenti_fischer_05}, along with Yonsei-Yale isochrones \citep{demarque_04}. The models self-consistently incorporate the high-resolution stellar spectrum measured with HIRES, direct stellar radius measured with CHARA, and stellar luminosity determined from a SED. Results for the physical properties of HR7672A, including its radius, luminosity, effective temperature, metallicity, mass, and age are shown in Table~\ref{tab:star_props}. Our mass estimate is confirmed using a separate analysis that is model independent and relies only upon the empirical mass-radius relations from \citet{torres_10}.

\begin{figure}[!t]								
\centering
\includegraphics[height=2.3in]{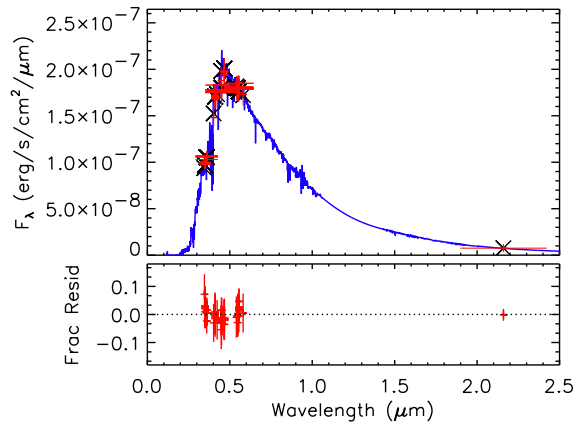} \\
\caption{Spectral energy distribution fit for HR7672 ($\chi^2_r=0.17$). The (blue) spectrum is a G0V spectral template \citep{pic98}. The (red) crosses indicate photometry values from the literature. ``Error bars'' in the horizontal direction represent bandwidths of the filters used. The (black) X-shaped symbols show the flux value of the spectral template integrated over the filter transmission. The lower panel displays the residuals around the fit in fractional flux units of photometric uncertainty.}
\label{fig:sed}
\end{figure}

\subsubsection{Stellar Diameter from Interferometry}\label{sec:diameter}
HR7672 was observed on the nights of September 18, 19, 2010, August 21, 2011, and October 2,3, 2011 with CHARA at Mount Wilson. We used the CHARA classic beam combiner in $H$-band ($\lambda_{central} = 1.67$~$\mu$m) with the longest available baselines, S1E1 (maximum 330~m) and E1W1 (maximum 313~m). Projected baseline lengths ranged between 245~m and 325~m during our observations. 

Our observing strategy is analogous to that employed in \citet{von11} and \citet{von11c}. We briefly repeat the principle components below. Our interferometric measurements include the common technique of taking bracketed sequences of the object with calibrator stars to characterize and eliminate effects from temporal variations of the atmosphere, telescope, and instrument upon our calculation of interferometric visibilities. We alternate between two nearby calibrators, HD187923 and HD192425, during the observations to minimize systematic errors.

The uniform disk, $\theta_{\rm UD}$, and limb-darkening corrected, $\theta_{\rm LD}$, diameters\footnote{The limb-darkening corrected $\theta_{\rm LD}$ corresponds to the angular diameter of the Rosseland, or mean, radiating surface of the star.} are found by fitting our calibrated visibility measurements to the respective functions for each relation. Specifically, we use a linear combination of Bessel functions \citep{han74,boyajian_09}. Limb darkening coefficients were taken from \citet{cla00}. The data and fit for $\theta_{\rm LD}$ are shown in Fig. \ref{fig:visibility}. The key to understanding the uniqueness of the solution is to recognize that the visibility must approach unity as the baseline approaches zero, and that degeneracies in the slope are broken by the absolute visibility value (as opposed to relative value) for a given baseline, so long as the location along the visibility curve is not mistaken for side-lobes in the Bessel function (which effectively corresponds to knowing the distance and luminosity class of the star). Our interferometric measurements yield uniform disk and limb-darkening corrected angular diameters of $\theta_{\rm UD} = 0.567 \pm 0.010$ mas and $\theta_{\rm LD} = 0.584 \pm 0.010$ mas respectively. Combined with the {\it Hipparcos} trigonometric parallax value of $\pi=56.28\pm0.35$ mas \citep{van_leeuwen_07}, we calculate the linear radius of HR7672A to be $R = 1.115\pm0.021R_{\rm \odot}$ (Table \ref{tab:star_props}). 


\subsubsection{Stellar Luminosity and Effective Temperature} \label{sec:temp}
Following the procedure outlined in \citet{vcb07}, we produce a fit to HR7672's SED based on the spectral templates of \citet{pic98} to literature photometry published in the references shown in Table \ref{tab:sedmeas}. A G0V spectral-type provides the best fitting template. Interstellar extinction is a free parameter in the fitting process and is calculated to be $A_V = 0.077 \pm 0.017$ mag. The SED fit for HR7672 along with its residuals are shown in Fig. \ref{fig:sed}. We calculate HR7672's stellar bolometric flux to be $F_{\rm BOL} = (1.360 \pm 0.028)\times10^{-7}$~erg cm$^{-2}$ s$^{-1}$, and its luminosity to be $L = 1.338 \pm 0.032 L_{\odot}$. Using a rewritten version of the Stefan-Boltzmann Law,
\begin{equation} \label{eq:temperature}
T_{\mbox{\small{eff}}} ({\rm K}) = 2341 (F_{\rm BOL}/\theta_{\rm LD}^2)^{\frac{1}{4}},
\end{equation}
where $F_{\rm BOL}$ is in units of $10^{-8}$~erg cm$^{-2}$ s$^{-1}$ and $\theta_{\rm LD}$ is in units of mas, we find HR7672 has an effective temperature of $T_{\mbox{\small{eff}}} = 5883 \pm 59$ K. 

\begin{figure}[!t]	
\begin{center}						
\includegraphics[height=2.3in]{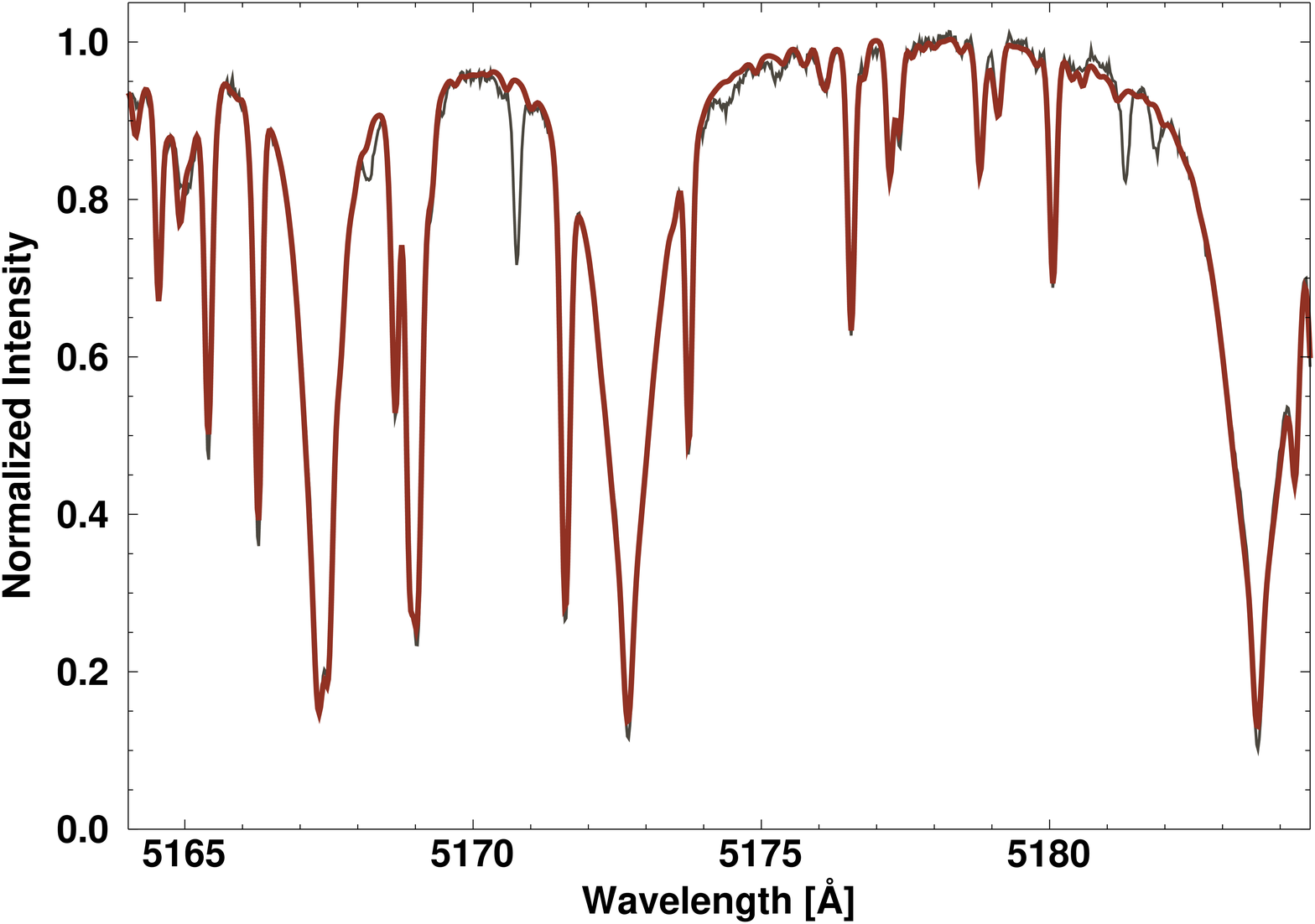} 
\end{center}
\caption{Comparison between a HIRES spectrum (black) and our best-fit SME model (maroon). We use these results along with Yonsei-Yale isochrones to derive the physical properties of HR7672A, self-consistently taking into account independent measurements of the stellar radius from interferometry and luminosity from an SED.}
\label{fig:sme}
\end{figure}

\subsubsection{Stellar Surface Gravity, Metallicity, Mass and Age}\label{sec:age}
Informed by the above results, we use high resolution spectroscopy, spectral modeling, and stellar isochrones to determine the remaining physical properties of the host star. A template spectrum of HR7672 was obtained with Keck/HIRES ($\lambda/\Delta \lambda = 55,000$) on Sept. 8, 2008. The iodine gas cell was removed from the optical path to isolate the absorption lines of the star. With an integration time of 17 seconds, we achieved a signal to noise ratio of 280 per pixel at 550 nm. 

We analyzed the stellar spectrum using the LTE spectral synthesis code {\it Spectroscopy Made Easy} (SME) described in \citet{valenti_fischer_05}. Recent versions of SME include an iterative scheme that ensures consistency between the surface gravity derived from synthetic spectra and evolutionary models \citep{valenti_09}. The stellar spectrum is fit using an initial guess for the surface gravity while the effective temperature, metallicity, macroscopic turbulence, and projected rotational velocity are varied to minimize residuals. With each iteration, results are compared to Yonsei-Yale isochrones \citep{demarque_04}. Model grids are interpolated to yield an estimate for the stellar mass, radius, age, and composition (see Fig. 1 of \citealt{valenti_09}). The spectral fitting procedure is repeated using the new (isochronal) surface gravity estimate in the following iteration until the procedure converges -- i.e., until the surface gravities agree. 

In the case of HR 7672, we fit the stellar spectrum by holding $T_{\small{eff}}$ fixed, setting it equal to discrete values within the allowable range derived from the Stefan-Boltzmann relation (see $\S$\ref{sec:temp}). Specifically, we analyzed the results for three separate cases: $T_{\small{eff}}=T_0$, $T_{\small{eff}}=T_0+\Delta T$, and $T_{\small{eff}}=T_0 - \Delta T$, where $T_0=5883$ K and $\Delta T = 59$ K ($\S$\ref{sec:temp}). The remaining parameters were varied as usual.\footnote{Alternatively, we could set the luminosity and radius input to SME equal to our measured values. By allowing the luminosity and radius to vary, we retain covariance between the parameters and demonstrate that the models naturally reproduce measured values while also accounting for uncertainty in the measurements.} We find that the $T_{\small{eff}}=T_0+\Delta T$ case provides the best fit to the HIRES spectrum. It is also the only solution for which the luminosity and radius match the directly measured values to within $1\sigma$ (the other cases are consistent to within 2$\sigma$). 

With this technique, we find a stellar surface gravity, $\mbox{log g}=4.42\pm0.06 \mbox{cm/s}^2$, metallicity, $\mbox{[Fe/H]}=0.05\pm0.07$, mass, $M_*=1.08\pm0.04M_{\odot}$, and age, $t_{\mbox{\small{age}}}=2.5\pm1.8$ Gyr. Fig.~\ref{fig:sme} shows a comparison between the HIRES spectrum and our best synthetic model. HR7672's physical properties are listed in (Tab.~\ref{tab:star_props}). Uncertainty in the stellar mass includes propagation of errors from the uncertainty in $T_{\small{eff}}$. We adopt this stellar mass for all dynamical calculations. 

To further validate our estimate of the host star mass, we use a separate technique that is independent of theoretical spectral and evolutionary models. \citet{torres_10} have compiled results for dozens of eclipsing binary stars and have constructed a polynomial that relates their measured masses and radii empirically. Using this relation along with the other parameters of HR7672A listed in Table~\ref{tab:star_props}, we find that HR7672A has a mass of $M_*=1.10\pm0.04M_{\odot}$. This result is in agreement with our analysis based on SME and Yonsei-Yale isochrones. Further, the \citet{torres_10} relations contain a known bias that systematically over-estimates near-solar ($M_*\approx1.0M_{\odot}$) stellar masses by as much as $\approx5\%$. Taking this into account indicates that our above result of $M_*=1.08\pm0.04M_{\odot}$ is accurate. Finally, we note that our age calculation affirms the original 1-3 Gyr estimate from \citet{liu_02}, which is based on an independent analysis using six different diagnostics, including: comparison of absolute visual magnitude to the location of the main-sequence, rotation period, X-ray emission, Ca II H+K emission, kinematics, and lithium absorption.   \\

\section{JOINT-FIT ORBITAL ANALYSIS}\label{sec:bayes}
We use Bayesian inference and Markov Chain Monte Carlo (MCMC) simulations to calculate the companion true mass, all six orbital elements, and their uncertainties. The RV and astrometry data are simultaneously fit with a Keplerian orbit. Stellar RV variations, $V_m(t_i)$, are modeled according to:
\begin{equation}
V_m(t_j)=K \left[\cos(\omega+f(t_j)) + e \: \cos(\omega) \right],
\end{equation}\label{equ:doppler}
where,
\begin{equation}
K=\left( \frac{2 \pi G}{P} \right)^{1/3}\frac{m \: \sin(i)}{\sqrt{1-e^2}(M_*+m)^{2/3}},
\end{equation}
is the RV semi-amplitude, $\omega$ is the argument of periastron, $f(t_j)$ is the true anomaly at epoch $t_j$, $e$ is the eccentricity, $i$ is the inclination, $P$ is the orbital period, $M_*$ is the mass of the star, and $m$ is the mass of the companion. Astrometry operates in the orthogonal direction and relates the semimajor axis, $a$, to the orbital period and system mass through Kepler's equation,
\begin{equation}\label{equ:kepler}
P^2=\frac{4 \pi^2 a^3}{G (M_*+m)},
\end{equation}
such that the orbit corresponds to the sky-projected separation of the companion at a given epoch (see Equ.~\ref{equ:likelihood}) and system parallax. 

The following physical parameters are used as variables in the analysis: the companion period ($P$), eccentricity ($e$), inclination ($i$), argument of periastron ($\omega$), longitude of the ascending node ($\Omega$), time of periastron passage ($t_p$), and true mass ($m$). Several nuisance parameters are also required, including differential RV offsets between HIRES and the Hamilton echelle at Lick, to account for the different instrument settings, as well as RV ``jitter" to account for astrophysical noise and ensure that the data are weighted properly \citep{isaacson_fischer_10}. RV jitter was added directly to the measurement uncertainties, which results in a reduced chi-squared statistic, $\sqrt{\chi_r^2}$, near unity. Given the long time baseline of the observations, a total of six different instrument offsets and seven different jitter terms were required (five from Lick and two from Keck; see Tab.~\ref{tab:offsets}). As expected, the measurement uncertainties decrease with time as well as the relative size of the required jitter, meaning that the RV precision and our ability to estimate the RV precision has improved significantly since September 1987. 

\begin{figure*}[!t]
\begin{center}
\includegraphics[height=2.5in]{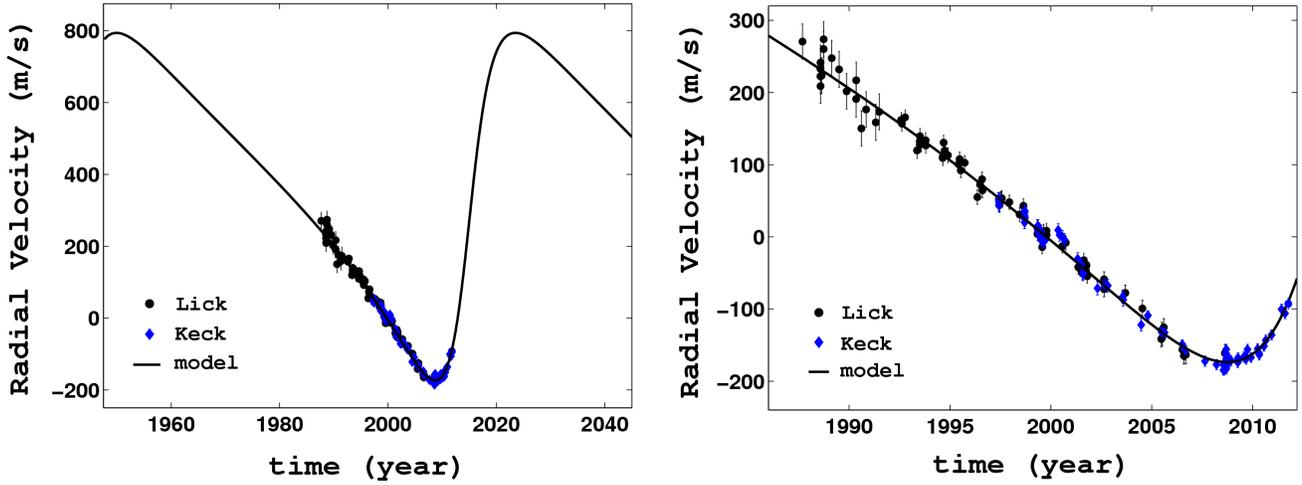} 
\caption{RV data and best-fit model. The Doppler signal shows significant curvature indicative of a high eccentricity. (left) RV time series showing a full predicted orbit cycle. (right) A closer view of the 24 year data set.} 
\end{center}
\end{figure*}

\begin{figure*}[!t]
\begin{center}
\vspace{0.4in}
\includegraphics[height=3.0in]{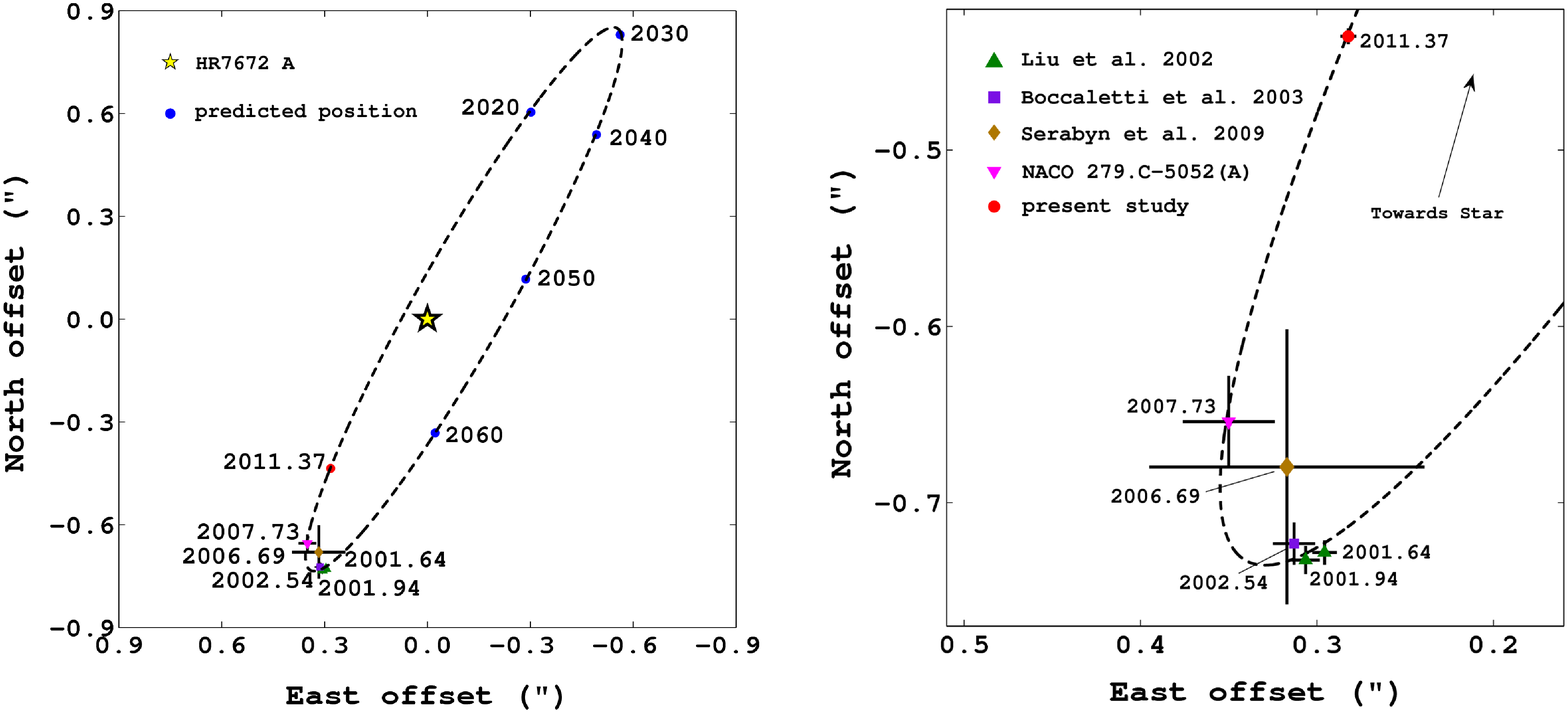} 
\caption{The orbit of HR7672B. (left) Full orbit showing recent astrometric measurements (south-east), the best-fit model, and our predictions for where the companion will be located at future epochs (blue dots -- dates correspond to Sept. 1). (right) A closer view of the south-east quadrant showing measurements from \citet{liu_02}, \citet{boccaletti_03}, \citet{serabyn_09}, NACO archival data, and our most recent measurement using NIRC2 at Keck, respectively. A much larger plate scale was used for the \citet{serabyn_09} observations compared to the other studies. The best-fit model passes through the ($1\sigma$) errorbar of each measurement.} 
\end{center}
\end{figure*} 

We use the Metropolis-Hastings algorithm to efficiently explore the multi-dimensional parameter space. A likelihood function, $L$, self-consistently relates the data, model parameters, and prior information, to the posterior probability density distribution \citep{ford_06,johnson_11_18pack}. The likelihood function used for calculations is given by:
\begin{small}
\begin{eqnarray}\label{equ:likelihood}
&&  \ln L  = -\sum_{j=1}^{N_{RV}} \ln \sqrt{2\pi(\sigma_j + s_{\ell})^2} - \frac{1}{2}\sum_{j=1}^{N_{RV}} \left[ \frac{\Delta V(t_j)}{\sigma_j+s_{\ell}} \right]^2   \nonumber  \\
&& - \sum_{k=1}^{N_{Ast}} \left[ \ln \sqrt{2\pi \sigma_{X_k}^2} + \ln\sqrt{2\pi \sigma_{Y_k}^2}\: \right] \nonumber   \\
&&  - \frac{1}{2}\sum_{k=1}^{N_{Ast}} \left[ \frac{\Delta X(t_k)^2} {\sigma_{X_k}^2} + \frac{\Delta Y(t_k)^2} {\sigma_{Y_j}^2} \right],    
\end{eqnarray}
\end{small}
\\ 
{\noindent}where $\Delta V(t_j)=V_j-V_m(t_j)$ represents the difference between the RV data and the Keplerian model RV at epoch $t_j$; $\Delta X(t_k)$ and $\Delta Y(t_k)$ represent the difference in the east and north position of the companion between the astrometric data and the Keplerian model astrometry at epoch $t_k$ respectively; $\sigma_i$ and $\sigma_{X_k}$, $\sigma_{Y_k}$ are the individual RV and astrometric uncertainties; and $s_{\ell}$ is the RV jitter for instrument setting $\ell$ (summation over the various instrument settings is implicit). We assume that both the RV and astrometric measurements follow Gaussian distributions. 

We randomly select one parameter per MCMC iteration to alter. Candidate transition probability functions follow a normal distribution with adjustable width and are centered on the most recently accepted parameter value. We do not change the transition function widths following the ``burn-in" stage. The algorithm accounts for the covariance between $\omega$ and $t_p$ by taking steps in $\omega \pm f_0$, where $f_0$ is the true anomaly of the companion at the first RV observing epoch. This approach improves the MCMC acceptance rate and accelerates convergence \citep{ford_06}. We use uniform priors for each parameter. The range of values is, however, truncated to reasonable limits. For example, we only consider companion masses in the range $20 \leq m / M_J \leq 120$. The eccentricity and other orbit parameters span the full range of possible values. 

To identify the global minimum, we compare the results of multiple chains that explore the likelihood manifold starting from different initial states. Convergence is reached once the Gelman-Rubin statistical criterion is met (Gelman \& Rubin 1992). Specifically, we require that:
\begin{equation}
R(z)=\sqrt{\frac{\bar{S}(z)+M(z)}{\bar{S}(z)}} \leq 1.1,
\end{equation} 
for each parameter $z$, where $\bar{S}(z)$ is the mean of the variance of the MCMC chains, and $M(z)$ is the variance of the mean of the MCMC chains (see \citealt{ford_06} for details). Following convergence, the individual chains are combined (linked together) to create the final parameter distributions. We find that $\approx10^8$ iterations are required to provide a sufficiently dense sampling of the posterior distribution.  


Uncertainty in the distance to HR7672 is self-consistently folded into the analysis by drawing random distance values from a normal distribution centered on the {\it Hipparcos} result for each MCMC iteration. The width of the distribution is set to match the measurement error. We account for uncertainty in the mass of the host star in the same way as the uncertainty in the distance, by drawing random stellar mass values from a normal distribution centered on the stellar mass estimate ($\S$\ref{sec:age}). The results are then combined with the MCMC chains in accordance with the above equations. This technique of incorporating uncertainty in the distance to HR7672 and the mass of the primary is robust because it preserves the covariance between orbital parameters captured by the MCMC calculations.

\section{RESULTS}\label{sec:results}
The RV and astrometry data used in combination with the distance measurement and stellar mass estimate allow the MCMC simulations to consistently converge for each parameter. Only a narrow range of orbital solutions exist that can satisfy the constraints imposed by both data sets simultaneously. Figures 5 and 6 display the final RV and astrometry data along with the best-fit model. The orbit well-matches the RV observations and also falls within the ($1\sigma$) uncertainty of each astrometric measurement. We find a reduced chi-square, $\chi_r$, value of $\sqrt{\chi_r^2}=0.96$ per degree of freedom overall. Fig. 7 shows the RV residuals following subtraction of our best-fit model from the data.

The final MCMC distributions are shown in Fig. 8 and the physical properties of HR7672B are summarized in Table~\ref{tab:results}. We find that HR7672B has a high eccentricity, $e=0.50^{+0.01}_{-0.01}$, and a near edge-on orbit, $i=97.3^{+0.4}_{-0.5}$ deg. It reached greatest elongation (south-east portion of orbit) just following the \citet{liu_02} direct imaging discovery. The velocity of HR7672B is now increasing sharply as it approaches periastron. Our predictions for its location on Sept. 1 in the years 2020, 2030, 2040, 2050, and 2060 are shown in Fig. 6. With an impact parameter of 67 mas, the companion will soon disappear behind the host star, much like the extrasolar planet Beta Pictoris b has evaded detection in the past \citep{fitzgerald_09,lagrange_10}.  

\begin{figure}[!t]
\begin{center}
\includegraphics[height=2.2in]{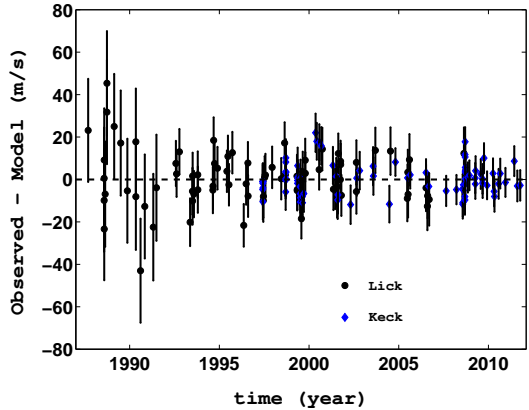} 
\caption{Doppler RV residuals. Precision RV techniques have improved dramatically over the past two decades. In addition to a long time baseline, HR7672 also has many samples per time because it is part of the NASA-UC $\eta$-Earth program. Some residual periodicities are noticeable but currently insufficient to claim evidence for the existence of an additional low-mass body with a comparatively short orbital period.} 
\end{center}\label{fig:residuals}
\end{figure} 

\begin{table*}
\centerline{
\begin{tabular}{ccc}
\hline
\hline
HR7672A  & Value & Technique \\
\hline 
$\theta_{\rm LD}$ (mas)	\dotfill	                       &	$0.584 \pm 0.010$			& 	interferometry                        \\
Radius ($R_{\rm \odot}$) \dotfill	                       &	$1.115 \pm 0.021$			&	interferometry, parallax          \\
Luminosity ($L_{\rm \odot}$) \dotfill	                &   $1.338 \pm 0.032$	                 &	spectral energy distribution    \\
$T_{\mbox{\small{eff}}}$ (K)	\dotfill	    	         &   $5883 \pm 59$           		      & 	Stefan-Boltzmann relation      \\
$\log[R'_{HK}]$ \dotfill                                                 &   $-4.854\pm0.025$                        &    HIRES spectroscopy \\
$V_{\mbox{\small{macro}}}$ (km/s) \dotfill       &    $4.8$                                &   iterative SME  \\
$V\sin i$ (km/s)              \dotfill                            &   $2.1\pm0.7$                      &  iterative SME  \\
$\mbox{[Fe/H]}$      \dotfill                                  &   $0.05 \pm 0.07$                        &   iterative SME   \\
log g  $(\mbox{cm/s}^2)$      \dotfill                   &   $4.42 \pm 0.06$                         &   iterative SME   \\
Mass ($M_{\rm \odot}$) \dotfill		                & 	$1.08 \pm 0.04$			       &   iterative SME    \\
Age (Gyr)	\dotfill			                          &    $2.5 \pm 1.8$		      	       & 	 iterative SME  \\
\hline
\hline
\end{tabular}}
\caption{The physical properties of HR7672A. We measure the stellar radius directly using CHARA interferometry. The luminosity is found by fitting a spectral energy distribution from photometry obtained over a broad wavelength range. The effective temperature derived from the Stefan-Boltzmann equation is held fixed as input to SME (sampled separately at $T_{eff}=5883\pm59$ K). Several SME iterations result in convergence of the stellar surface gravity (log g) between the spectral model and Yonsei-Yale isochrones. The final estimated mass, age, and radius are consistent with the measured stellar luminosity and radius to within $1\sigma$. Our derived stellar mass is also consistent with the model-independent empirical relations of \citet{torres_10}. Our derived stellar age is consistent with the six different age diagnostics of \citet{liu_02}. HR7672A is a G0V star and thus a convenient calibrator. Its physical properties may be used to infer those of HR7672B, such as metal content and age.} 
\label{tab:star_props}
\end{table*}

\begin{table*}[!t]
\centerline{
\begin{tabular}{cccc}
\hline
\hline
HR7672B                                         &  weighted mean           &  68.2\% CI            &  95.4\% CI       \\
\hline
\hline
mass ($M_J$)  \dotfill                      &     68.7                  &      65.6--71.1     &   63.2--74.6    \\  
P (yr)                \dotfill                       &    73.3                  &     70.4--75.5       &    68.5--79.0   \\
a (AU)              \dotfill                        &    18.3                 &     17.8--18.7       &    17.5--19.3  \\
$e$                               \dotfill            &    0.50              &  0.49--0.51     &  0.48--0.52  \\
$i$ $(\mbox{deg})$       \dotfill           &     97.3               &   96.8--97.7        &  96.4--98.1      \\
$\omega$ $(\mbox{deg})$  \dotfill     &    259                  &   257--261          &  254--263        \\
$\Omega$ $(\mbox{deg})$  \dotfill    &    61.0                 &   60.6--61.3         & 60.2--61.6      \\
$t_p$ (year)                \dotfill             &    2014.6           &  2014.5--2014.7    &   2014.3--2014.8   \\
\hline
\hline
\end{tabular}}
\caption{Companion dynamical mass and three-dimensional orbit parameters resulting from our MCMC analysis. Both 68.2\% and 95.4\% confidence intervals (CI) are listed. HR7672B is a benchmark brown dwarf with high eccentricity.}
\label{tab:results}
\end{table*}

\begin{table}[!t]
\centerline{
\begin{tabular}{ccc}
\hline
\hline
Filter                     &       HR~7672 A           &   HR~7672 B         \\
\hline
\hline
V                           &         5.80                        &       ---                          \\
J                            &         4.69                       &    $\approx14.39$     \\
H                           &         4.43                       &    $14.04\pm0.14$      \\
K                           &         4.49                       &    $K_s=13.04\pm0.10$    \\
\hline
\hline
\end{tabular}}
\caption{Apparent magnitudes of the HR7672 system. Values for HR~7672 A are from SIMBAD. Values for HR~7672 B are from \citet{boccaletti_03}.}
\label{tab:mags}
\end{table}

\begin{figure*}[!t]
\begin{center}
\includegraphics[height=3.35in]{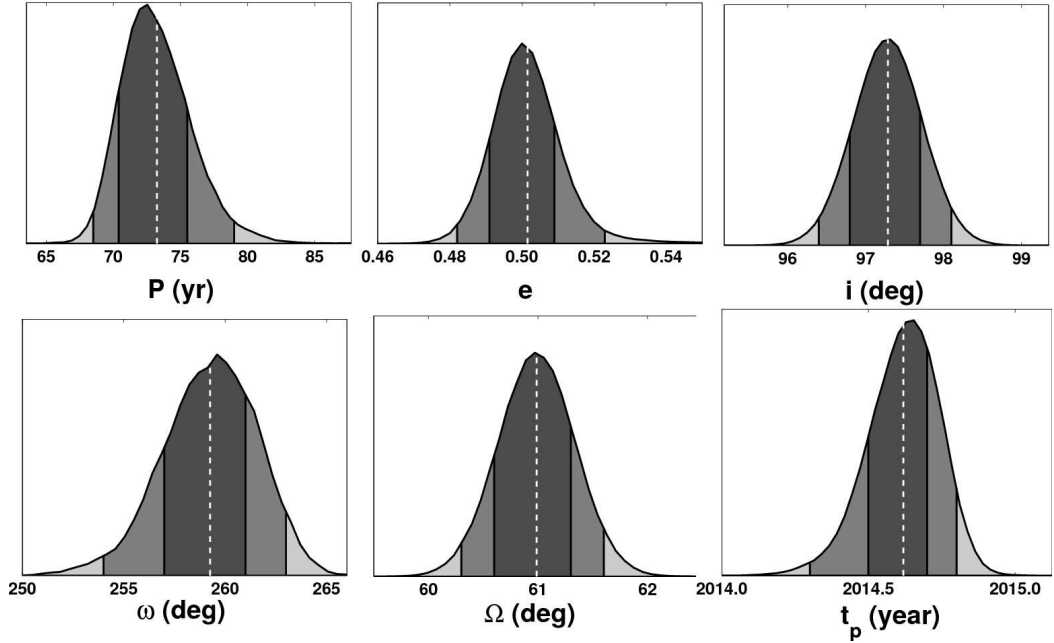} 
\caption{Results from MCMC calculations showing the marginalized posterior distributions of the companion period ($P$), eccentricity ($e$), inclination ($i$), argument of periastron ($\omega$), longitude of the ascending node ($\Omega$), and time of periastron passage ($t_p$). A vertical dashed line indicates the weighted mean value for each parameter. Shaded regions correspond to 68.2\% and 95.4\% confidence intervals respectively.} 
\end{center}\label{fig:dists}
\end{figure*} 

\begin{figure*}[!t]
\begin{center}
\includegraphics[height=2.9in]{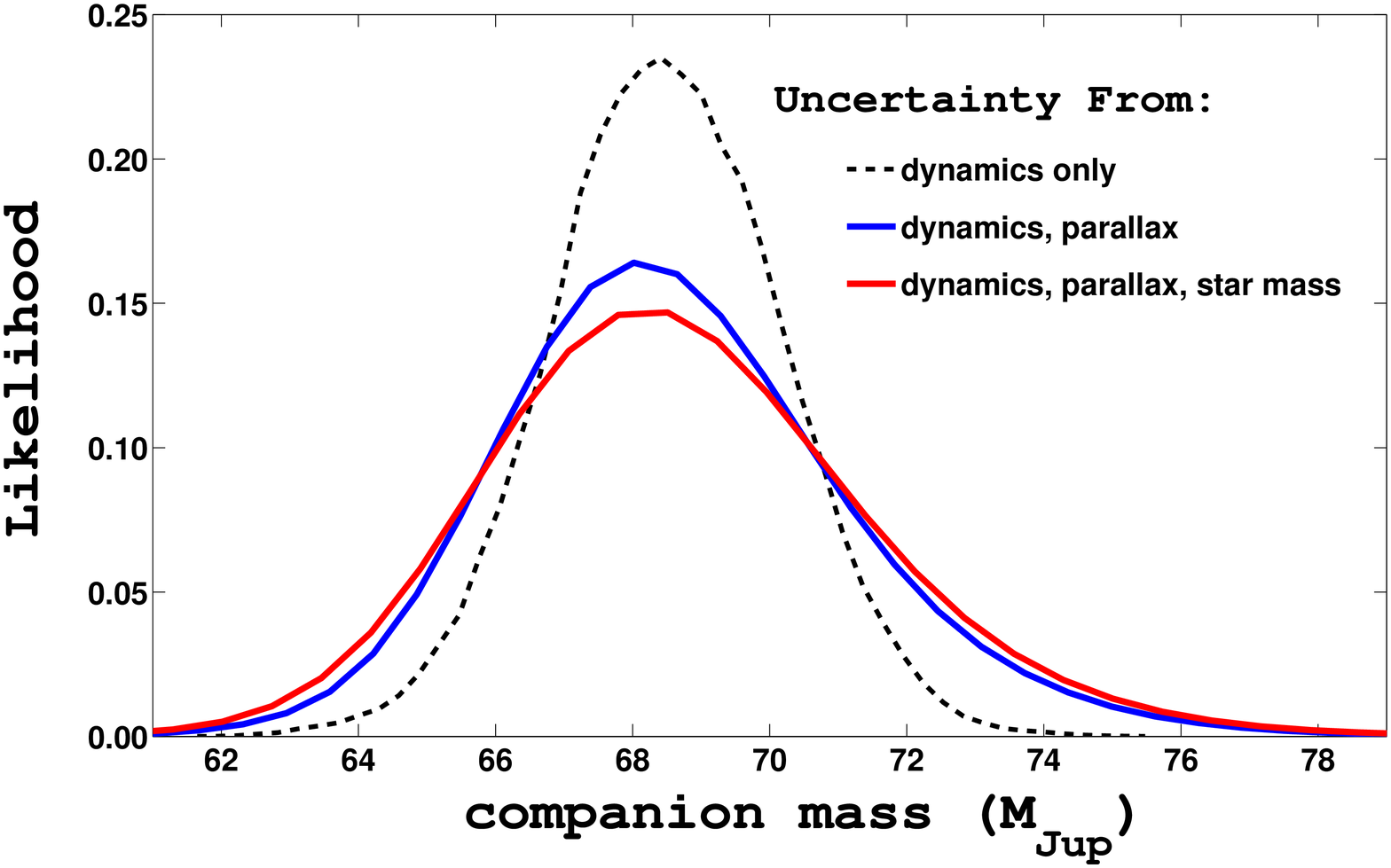} 
\caption{Companion posterior mass distribution shown before and after taking into account uncertainty in the system parallax and mass of the primary star. Uncertainty in the companion mass is limited primarily by having only partial orbital phase coverage.} 
\end{center}\label{fig:mp_dist}
\end{figure*} 

\begin{figure*}[!t]
\begin{center}
\includegraphics[height=3.4in]{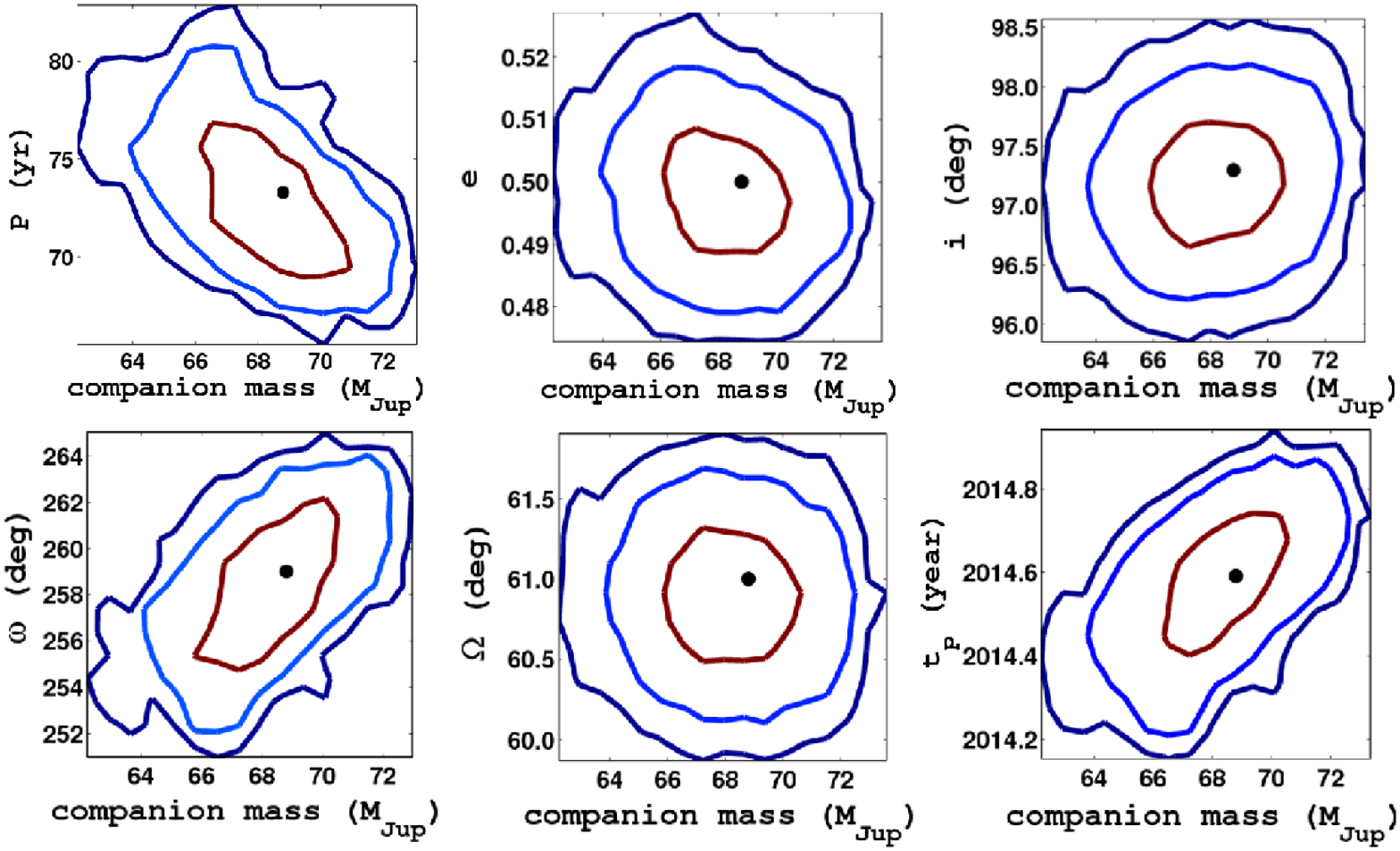} 
\caption{Probability matrices showing the covariance between the companion mass and all six orbital elements. Each panel shows the posterior distribution of two parameters marginalized over the remaining variables including RV offsets and jitter. The contours indicate iso-probability levels corresponding to 68.2\% (red), 95.4\% (light blue), and 99.7\% (dark blue) respectively, prior to accounting for uncertainty in the system parallax and star mass. A black circle denotes the best-fit values. The companion mass is strongly covariant with the orbital period and argument of periastron (and likewise time of periastron passage).} 
\end{center}\label{fig:covariance}
\end{figure*} 

We find that HR7672B has a mass of $m=68.7^{+2.4}_{-3.1}M_J$ and thus resides in an interesting regime that borders, but lies just beneath, the hydrogen-fusing limit (Fig. 9). Only 9.9\% of the distribution lies above the canonical value of $\approx$$72M_J$ often taken as the dividing line between brown dwarfs and stars \citep{chabrier_00}. We therefore conclude that HR7672B is a brown dwarf at 90.1\% confidence (Fig. 11), assuming the companion has the same near-solar metallicity as the primary, which we measure to be [Fe/H]$=0.05\pm0.07$ dex. 

Figure 9 shows the companion mass distribution before and after taking into account the uncertainty in parallax and stellar mass. Uncertainty in the companion mass is limited primarily by having only a partial orbit, motivating the need for continued Doppler and astrometric monitoring. Uncertainty in parallax is the second largest effect. Significant improvements in parallax over the current {\it Hipparcos} measurement will likely require dedicated observations from space. Uncertainty in the mass of the primary star is a comparatively small effect, since the companion mass depends weakly upon star mass with Doppler observations. For example, an error of $\pm0.13M_{\odot}$ in the host star mass would be required to shift the companion mass exterior to its current 95.4\% confidence interval. Covariance matrices between the six orbit parameters and companion mass are shown in Fig. 10. 

\begin{figure}[!t]
\begin{center}
\includegraphics[height=2.25in]{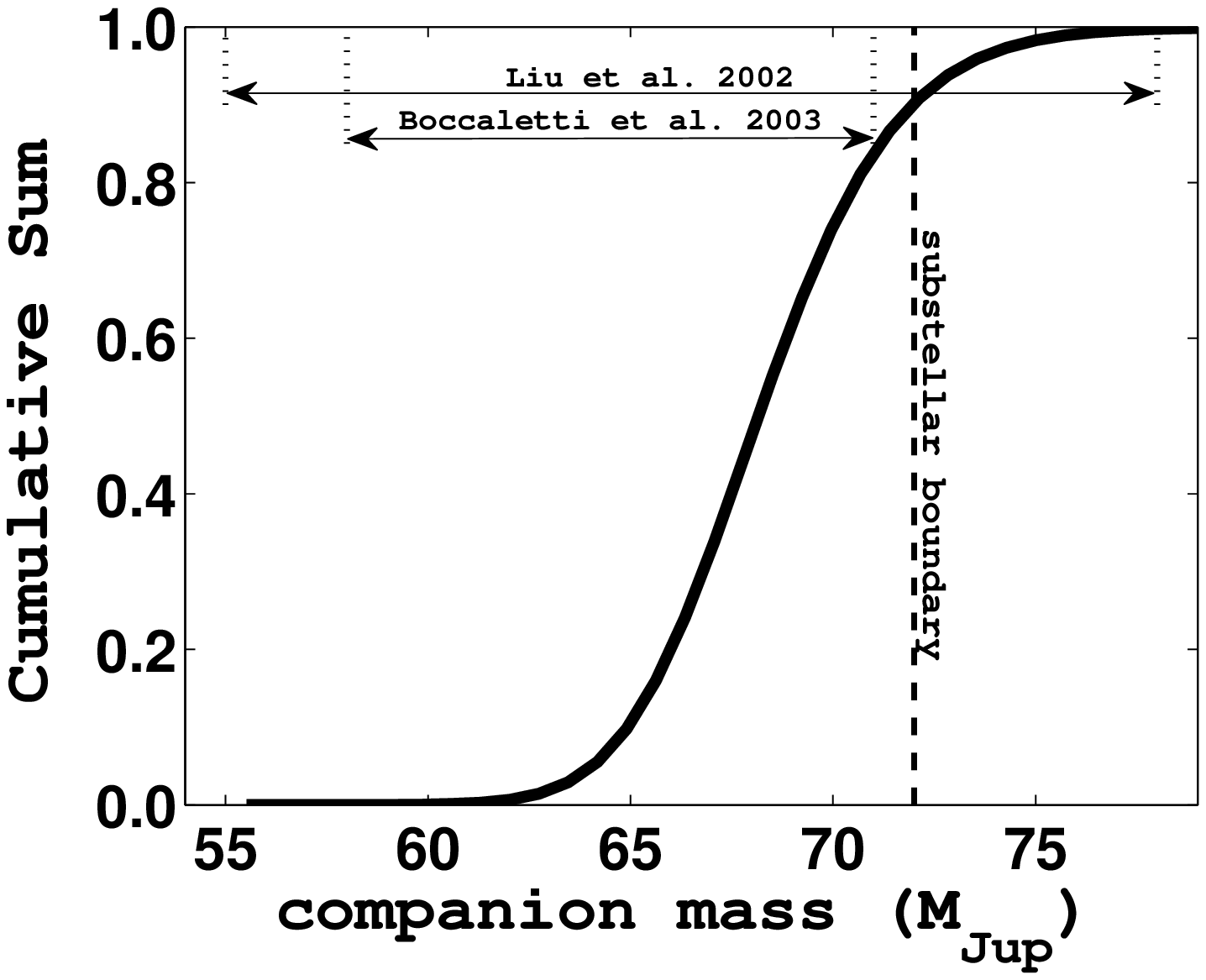} 
\caption{Cumulative summation of the posterior companion mass distribution. The canonical boundary at $72M_J$ separating stars from brown dwarfs is labeled for reference. HR7672B is a brown dwarf at 90.1\% confidence. Results from orbital dynamics are more accurate and precise than estimates based on spectrophotometry.} 
\end{center}\label{fig:cumsum}
\end{figure}

The final companion mass distribution is narrow compared to the range of possible masses resulting from spectrophotometry. For example, \citet{liu_02} find that the companion has a mass between $55-78M_J$, using direct K-band spectroscopy, a number of age diagnostics, and the theoretical models of \citet{burrows_97} and \citet{chabrier_00}. \citet{boccaletti_03} have obtained complementary photometry in the J, H bands to refine this estimate. Using the same age range of 1-3 Gyr from \citet{liu_02}, they find that HR7672B has a mass between $58-71M_J$. Performing similar calculations with the \citet{baraffe_03} evolutionary models, we find that HR7672B should have a mass in the $57-74M_J$ range. The \citet{baraffe_03} models are in excellent agreement with the more recent \citet{saumon_08} models for L-dwarfs older than several Myr. Thus, in each case our calculations using orbital dynamics are already more precise and accurate than estimates based on analyzing light received directly from the companion (Fig. \ref{fig:cumsum}). 

\section{SUMMARY \& DISCUSSION} 
We present the first three-dimensional orbit solution and mass determination for a directly imaged L-dwarf companion to a solar-type star. HR7672 is a unique benchmark system because: (i) the primary is a G0V star amenable to ultra-precise Doppler measurements; (ii) the metallicity of the primary is reliably determined from spectroscopy; (iii) the mass and age of the primary are well-constrained; and (iv) the parallax has been measured with a fractional error of only 1.2\%. 

We have monitored HR7672 with precise RV measurements for 24 years, covering 33\% of an orbital cycle. The observational cadence is high compared to other trend stars because this system was also a target of the NASA-UC $\eta$-Earth program \citep{howard_10}, which began conveniently around the same time as the Doppler signal started to show curvature. We have also obtained a recent high-contrast image of the system with NIRC2 at Keck, significantly increasing the astrometric time baseline by extending the sky-projected orbital coverage to 13\% of a full cycle. Both data sets have a high ($\approx$100) signal-to-noise ratio and exhibit substantial orbital motion. We have performed a joint Doppler-astrometric MCMC analysis to compute the companion orbital elements, true mass, and their uncertainties. 

We find that HR7672B has a near edge-on, $i=97.3^{+0.4}_{-0.5}$ deg, highly eccentric, $e=0.50^{+0.01}_{-0.01}$, orbit, with semimajor axis $a=18.3^{+0.4}_{-0.5}$ AU. Presently accelerating on its path back towards the primary star, both from the perspective of the Earth and from the perspective of HR7672A, this companion will gradually becoming more difficult to image directly. We predict that HR7672B will reach periastron in the year 2014 just prior to disappearing behind HR7672A. This companion does not yet have JH spectra. We recommend that high-contrast imaging programs employing integral-field spectrographs observe this target as soon as possible, while the components are reasonably well separated. 

The mass of HR7672B is $68.7^{+2.4}_{-3.1}M_J$. This L$4.5\pm1.5$-dwarf therefore resides near the stellar/substellar boundary, though is likely a brown dwarf if not a transition object: 90.1\% of the mass probability distribution falls below the $72M_J$ dividing line. Our dynamical measurement is more accurate and precise than that obtained previously from spectrophotometry and the use of theoretical models. With a fractional error of only 4\%, it is comparable to the results for transiting brown dwarfs \citep{stassun_06,johnson_11_lhs6343}. Uncertainty in the mass will decrease further as additional RV and astrometric measurements are obtained.  

These observations provide a mass determination that is completely independent of the companion spectrum, colors, and luminosity (see Appendix A for an estimate of the companion bolometric luminosity). Presently, few benchmark systems have dynamical mass measurements and provide such comprehensive information. For instance, GD165B has been well-characterized with optical and near-infrared spectroscopy, but its age is uncertain because it orbits a white dwarf \citep{kirkpatrick_99,jones_94}. HD203030B has also been studied across a wide bandpass, conveniently orbits a G8V star, and has a narrow age range (130-400 Myr), but its projected separation is 487 AU, thus prohibiting a dynamical mass \citep{metchev_hillenbrand_06}. Likewise, HD3651B and HD114762B orbit solar-type stars but their projected separations are 480 AU and 128 AU respectively \citep{liu_07,bowler_09}. Many other dynamical measurements have targeted companions to M-stars, thus helping to circumvent issues with contrast \citep{zapatero_04,ireland_08_BD,dupuy_09_lhs,konopacky_10}, but M-star metallicities are notoriously difficult to determine \citep{johnson_09}. Low-mass companions to solar-type stars have also been studied using Doppler measurements in combination with HST astrometry \citep{benedict_10} and Hipparcos intermediate astrometry \citep{sahlmann_11}, but indirect observations preclude a comparison between spectral models and dynamical mass measurements by definition. One comparable system, HD~130948BC, consists of a double brown dwarf orbiting a G2V primary. Although RVs have not been obtained for the pair, which are nominally separated by $\approx$100 mas, a near-unity luminosity ratio permits a precise individual mass determination \citep{dupuy_09,konopacky_10}. We conclude that HR7672B is thus a rare and precious mass, age, and metallicity benchmark brown dwarf companion that may be used to explicitly calibrate theoretical evolutionary models and synthetic spectral models by anchoring input parameters to measured values. 

Detailed substellar companion characterization studies are particularly relevant now that the first extrasolar planet spectra are securely in hand \citep{bowler_10,barman_11_2m1207,barman_11}. As the next generation of high-contrast imaging instruments come online (e.g., \citealt{hinkley_11_PASP}), and more cold bodies orbiting nearby stars are discovered, HR7672B will serve as a helpful guide for the development of more sophisticated theoretical models by providing a link between our understanding of stars and planets. 

\section{APPENDIX A: THE LUMINOSITY OF HR7672B}
We can estimate the luminosity of HR7672B using its apparent magnitude and parallax by applying a near-infrared bolometric correction derived from M, L, T field dwarfs. The apparent magnitudes of HR7672B are measured to be $m_J\approx14.39\pm0.20$, $m_H=14.04\pm0.14$, and $m_{K_s}=13.04\pm0.10$ \citep{boccaletti_03}. Of these, the K-band magnitude is regarded as most reliable, given the contrast ratio between the companion and host star. We recalculate the absolute magnitude from \citet{boccaletti_03}, owing to the revised Hipparcos parallax since the time of publication, finding $M_{K_s}=11.79\pm0.10$. \citet{liu_02} arrive at a similar answer of $M_{K_s}=11.8\pm0.1$ using relations between spectral type and absolute magnitudes from \citet{kirkpatrick_00}. 

To calculate a bolometric magnitude, we next derive an estimate for the bolometric correction using the empirical relations from \citet{golimowski_04}. Correcting for the expected change in magnitude between the $K_s$ filter and $K$ filter, which results in an $\approx0.11$ mag increase in brightness (i.e., \citet{rudy_96}), and using a spectral-type of L4.5, we find BC$_K=3.35\pm0.13$, where the uncertainty is given by the rms scatter in the polynomial fit. The bolometric magnitude is thus
\begin{equation}
M_{BOL}=M_K+BC_K=15.05\pm0.23.
\end{equation}
Using the Sun as a reference with $M_{BOL,\odot}=4.74$, we find the luminosity of HR7672B is equal to $L=(7.5\pm1.6)\times10^{-5}L_{\odot}$. With an age of $t_{\mbox{\small{age}}}=2.5\pm1.8$ Gyr, this result appears to be consistent with theoretical cooling models \citep{burrows_97}. Applying the Stefan-Boltzmann equation, we estimate the radius to be $R\approx1.0\pm0.4R_J$ assuming an effective temperature in the middle of the $1510-1850$ K range found by \citet{liu_02}. 

\section{APPENDIX B: THE SPIN AXIS OF HR7672A}
We can also estimate the inclination of the primary star rotation axis. Modeling of HR7672A's spectrum yields a projected rotational velocity of $V \mbox{sin} i = 2.1\pm0.7$ km/s. Using our measurement of $\log{R'_{HK}}=-4.854\pm0.025$ for the chromospheric Ca II H and K emission, we employ the \citet{noyes_84} empirical relations to estimate the rotational period, $P_{\mbox{\small{rot}}}$. With $B-V=0.61$ and a convective turnover time of $\log ( \tau / \mbox{days})=0.99\pm0.06$, we find $P_{\mbox{\small{rot}}}=17.5\pm2.3$ days, assuming a mixing length to scale height ratio of 1.9 (see \citealt{wright_04}, \citealt{noyes_84}, and references therein for details). Combining this result with our direct radius measurement ($\S$\ref{sec:diameter}), we find that the inclination of the stellar rotation axis is $i_{\mbox{\small{rot}}}=39^{\circ} \pm 16^{\circ}$, where $i_{\mbox{\small{rot}}}=0$ corresponds to a pole-on orientation. Thus, the spin axis of HR7672A and the orbit of HR7672B are not aligned. 

\section{Acknowledgements}
We thank Brendan Bowler for drawing our attention to the VLT archival images of HR7672 taken with NACO in September 2007, Dimitri Mawet for help interpreting the archival data, and Randy Campbell for assisting with observations during our May 2011 astrometric measurement with NIRC2. The anonymous referee, Trent Dupuy, and Brendan Bowler provided helpful suggestions that improved the quality of the paper. We would also like to thank R. Paul Butler, Chris McCarthy, and Steve Vogt for making many of the early RV observations.

We thank the University of California for the use of Lick Observatory, including both the Coude Auxiliary Telescope and the 3-meter. A large portion of the data presented herein were obtained at the W.M. Keck Observatory, which is operated as a scientific partnership among the California Institute of Technology, the University of California and the National Aeronautics and Space Administration. The Observatory was made possible by the generous financial support of the W.M. Keck Foundation. We acknowledge usage of the ESO scientific archive for providing complementary astrometric data from the VLT. This research made use of the SIMBAD literature database, operated at CDS, Strasbourg, France, and of NASA's Astrophysics Data System. This publication makes use of data products from the Two Micron All Sky Survey, which is a joint project of the University of Massachusetts and the Infrared Processing and Analysis Center/California Institute of Technology, funded by the National Aeronautics and Space Administration and the National Science Foundation. This research made use of the NASA/IPAC/NExScI Star and Exoplanet Database, which is operated by the Jet Propulsion Laboratory, California Institute of Technology, under contract with the National Aeronautics and Space Administration. TSB acknowledges support provided by NASA through Hubble Fellowship grant HST-HF-51252.01 awarded by the Space Telescope Science Institute, which is operated by the Association of Universities for Research in Astronomy, Inc., for NASA, under contract NAS 5-26555. The CHARA Array is funded by the National Science Foundation through NSF grants AST-0606958 and AST-0908253 and by Georgia State University through the College of Arts and Sciences, the W. M. Keck Foundation, the Packard Foundation, and the NASA Exoplanet Science Institute. 


\begin{small}
\bibliographystyle{jtb}
\bibliography{ms.bib}

\begin{thebibliography}{}

\bibitem[\protect\astroncite{{Absil} and {Mawet}}{2010}]{absil_mawet_10}
{Absil}, O. and {Mawet}, D. (2010) ,
\newblock {\em \aapr} {\bf 18}, 317

\bibitem[\protect\astroncite{{Aitken}}{1919}]{aitken_19}
{Aitken}, R.~G. (1919) ,
\newblock {\em \pasp} {\bf 31}, 196

\bibitem[\protect\astroncite{{Allard} et~al.}{1997}]{allard_97}
{Allard}, F., {Hauschildt}, P.~H., {Alexander}, D.~R., and {Starrfield}, S.
  (1997) ,
\newblock {\em \araa} {\bf 35}, 137

\bibitem[\protect\astroncite{{Baraffe} et~al.}{2003}]{baraffe_03}
{Baraffe}, I., {Chabrier}, G., {Barman}, T.~S., {Allard}, F., and {Hauschildt},
  P.~H. (2003) ,
\newblock {\em \aap} {\bf 402}, 701

\bibitem[\protect\astroncite{{Barman} et~al.}{2011a}]{barman_11}
{Barman}, T.~S., {Macintosh}, B., {Konopacky}, Q.~M., and {Marois}, C. (2011a)
  ,
\newblock {\em \apj} {\bf 733}, 65

\bibitem[\protect\astroncite{{Barman} et~al.}{2011b}]{barman_11_2m1207}
{Barman}, T.~S., {Macintosh}, B., {Konopacky}, Q.~M., and {Marois}, C. (2011b)
  ,
\newblock {\em ArXiv e-prints}

\bibitem[\protect\astroncite{{Barnes} et~al.}{1997}]{1997PASP..109..645B}
{Barnes}, III, T.~G., {Fernley}, J.~A., {Frueh}, M.~L., {Navas}, J.~G.,
  {Moffett}, T.~J., and {Skillen}, I. (1997) ,
\newblock {\em \pasp} {\bf 109}, 645

\bibitem[\protect\astroncite{{Basri} et~al.}{1996}]{basri_96}
{Basri}, G., {Marcy}, G.~W., and {Graham}, J.~R. (1996) ,
\newblock {\em \apj} {\bf 458}, 600

\bibitem[\protect\astroncite{{Becklin} and
  {Zuckerman}}{1988}]{becklin_zuckerman_88}
{Becklin}, E.~E. and {Zuckerman}, B. (1988) ,
\newblock {\em \nat} {\bf 336}, 656

\bibitem[\protect\astroncite{{Benedict} et~al.}{2010}]{benedict_10}
{Benedict}, G.~F., {McArthur}, B.~E., {Bean}, J.~L., {Barnes}, R., {Harrison},
  T.~E., {Hatzes}, A., {Martioli}, E., and {Nelan}, E.~P. (2010) ,
\newblock {\em \aj} {\bf 139}, 1844

\bibitem[\protect\astroncite{{Bernstein} and {Khushalani}}{2000}]{bernstein_00}
{Bernstein}, G. and {Khushalani}, B. (2000) ,
\newblock {\em \aj} {\bf 120}, 3323

\bibitem[\protect\astroncite{{Biller} et~al.}{2010}]{biller_10_pztel}
{Biller}, B.~A., {Liu}, M.~C., {Wahhaj}, Z., {Nielsen}, E.~L., {Close}, L.~M.,
  {Dupuy}, T.~J., {Hayward}, T.~L., {Burrows}, A., {Chun}, M., {Ftaclas}, C.,
  {Clarke}, F., {Hartung}, M., {Males}, J., {Reid}, I.~N., {Shkolnik}, E.~L.,
  {Skemer}, A., {Tecza}, M., {Thatte}, N., {Alencar}, S.~H.~P., {Artymowicz},
  P., {Boss}, A., {de Gouveia Dal Pino}, E., {Gregorio-Hetem}, J., {Ida}, S.,
  {Kuchner}, M.~J., {Lin}, D., and {Toomey}, D. (2010) ,
\newblock {\em \apjl} {\bf 720}, L82

\bibitem[\protect\astroncite{{Boccaletti} et~al.}{2003}]{boccaletti_03}
{Boccaletti}, A., {Chauvin}, G., {Lagrange}, A.-M., and {Marchis}, F. (2003) ,
\newblock {\em \aap} {\bf 410}, 283

\bibitem[\protect\astroncite{{Boden} et~al.}{2006}]{boden_06}
{Boden}, A.~F., {Torres}, G., and {Latham}, D.~W. (2006) ,
\newblock {\em \apj} {\bf 644}, 1193

\bibitem[\protect\astroncite{{Bowler} et~al.}{2010}]{bowler_10}
{Bowler}, B.~P., {Johnson}, J.~A., {Marcy}, G.~W., {Henry}, G.~W., {Peek},
  K.~M.~G., {Fischer}, D.~A., {Clubb}, K.~I., {Liu}, M.~C., {Reffert}, S.,
  {Schwab}, C., and {Lowe}, T.~B. (2010) ,
\newblock {\em \apj} {\bf 709}, 396

\bibitem[\protect\astroncite{{Bowler} et~al.}{2009}]{bowler_09}
{Bowler}, B.~P., {Liu}, M.~C., and {Cushing}, M.~C. (2009) ,
\newblock {\em \apj} {\bf 706}, 1114

\bibitem[\protect\astroncite{{Boyajian} et~al.}{2009}]{boyajian_09}
{Boyajian}, T.~S., {McAlister}, H.~A., {Cantrell}, J.~R., {Gies}, D.~R.,
  {Brummelaar}, T.~A.~t., {Farrington}, C., {Goldfinger}, P.~J., {Sturmann},
  L., {Sturmann}, J., {Turner}, N.~H., and {Ridgway}, S. (2009) ,
\newblock {\em \apj} {\bf 691}, 1243

\bibitem[\protect\astroncite{{Burrows} et~al.}{1997}]{burrows_97}
{Burrows}, A., {Marley}, M., {Hubbard}, W.~B., {Lunine}, J.~I., {Guillot}, T.,
  {Saumon}, D., {Freedman}, R., {Sudarsky}, D., and {Sharp}, C. (1997) ,
\newblock {\em \apj} {\bf 491}, 856

\bibitem[\protect\astroncite{{Burrows} et~al.}{2006}]{burrows_06}
{Burrows}, A., {Sudarsky}, D., and {Hubeny}, I. (2006) ,
\newblock {\em \apj} {\bf 640}, 1063

\bibitem[\protect\astroncite{{Butler} et~al.}{1996}]{butler_96}
{Butler}, R.~P., {Marcy}, G.~W., {Williams}, E., {McCarthy}, C., {Dosanjh}, P.,
  and {Vogt}, S.~S. (1996) ,
\newblock {\em \pasp} {\bf 108}, 500

\bibitem[\protect\astroncite{{Chabrier} et~al.}{2000}]{chabrier_00}
{Chabrier}, G., {Baraffe}, I., {Allard}, F., and {Hauschildt}, P. (2000) ,
\newblock {\em \apj} {\bf 542}, 464

\bibitem[\protect\astroncite{{Claret}}{2000}]{cla00}
{Claret}, A. (2000) ,
\newblock {\em \aap} {\bf 363}, 1081

\bibitem[\protect\astroncite{{Close} et~al.}{2005}]{close_05}
{Close}, L.~M., {Lenzen}, R., {Guirado}, J.~C., {Nielsen}, E.~L., {Mamajek},
  E.~E., {Brandner}, W., {Hartung}, M., {Lidman}, C., and {Biller}, B. (2005) ,
\newblock {\em \nat} {\bf 433}, 286

\bibitem[\protect\astroncite{{Cowley} et~al.}{1967}]{1967AJ.....72.1334C}
{Cowley}, A.~P., {Hiltner}, W.~A., and {Witt}, A.~N. (1967) ,
\newblock {\em \aj} {\bf 72}, 1334

\bibitem[\protect\astroncite{{Crawford} et~al.}{1966}]{1966AJ.....71..709C}
{Crawford}, D.~L., {Barnes}, J.~V., {Faure}, B.~Q., and {Golson}, J.~C. (1966)
  ,
\newblock {\em \aj} {\bf 71}, 709

\bibitem[\protect\astroncite{{Crepp} et~al.}{2011}]{crepp_11}
{Crepp}, J.~R., {Pueyo}, L., {Brenner}, D., {Oppenheimer}, B.~R., {Zimmerman},
  N., {Hinkley}, S., {Parry}, I., {King}, D., {Vasisht}, G., {Beichman}, C.,
  {Hillenbrand}, L., {Dekany}, R., {Shao}, M., {Burruss}, R., {Roberts}, L.~C.,
  {Bouchez}, A., {Roberts}, J., and {Soummer}, R. (2011) ,
\newblock {\em \apj} {\bf 729}, 132

\bibitem[\protect\astroncite{{Cumming} et~al.}{1999}]{cumming_99}
{Cumming}, A., {Marcy}, G.~W., and {Butler}, R.~P. (1999) ,
\newblock {\em \apj} {\bf 526}, 890

\bibitem[\protect\astroncite{{Currie} et~al.}{2011}]{currie_11_betapic}
{Currie}, T., {Thalmann}, C., {Matsumura}, S., {Madhusudhan}, N., {Burrows},
  A., and {Kuchner}, M. (2011) ,
\newblock {\em \apjl} {\bf 736}, L33+

\bibitem[\protect\astroncite{{Cushing} et~al.}{2011}]{cushing_11}
{Cushing}, M.~C., {Kirkpatrick}, J.~D., {Gelino}, C.~R., {Griffith}, R.~L.,
  {Skrutskie}, M.~F., {Mainzer}, A.~K., {Marsh}, K.~A., {Beichman}, C.~A.,
  {Burgasser}, A.~J., {Prato}, L.~A., {Simcoe}, R.~A., {Marley}, M.~S.,
  {Saumon}, D., {Freedman}, R.~S., {Eisenhardt}, P.~R., and {Wright}, E.~L.
  (2011) ,
\newblock {\em ArXiv e-prints}

\bibitem[\protect\astroncite{{Cushing} et~al.}{2008}]{cushing_08}
{Cushing}, M.~C., {Marley}, M.~S., {Saumon}, D., {Kelly}, B.~C., {Vacca},
  W.~D., {Rayner}, J.~T., {Freedman}, R.~S., {Lodders}, K., and {Roellig},
  T.~L. (2008) ,
\newblock {\em \apj} {\bf 678}, 1372

\bibitem[\protect\astroncite{{Cutri} et~al.}{2003}]{2mass_book}
{Cutri}, R.~M., {Skrutskie}, M.~F., {van Dyk}, S., {Beichman}, C.~A.,
  {Carpenter}, J.~M., {Chester}, T., {Cambresy}, L., {Evans}, T., {Fowler}, J.,
  {Gizis}, J., {Howard}, E., {Huchra}, J., {Jarrett}, T., {Kopan}, E.~L.,
  {Kirkpatrick}, J.~D., {Light}, R.~M., {Marsh}, K.~A., {McCallon}, H.,
  {Schneider}, S., {Stiening}, R., {Sykes}, M., {Weinberg}, M., {Wheaton},
  W.~A., {Wheelock}, S., and {Zacarias}, N. (2003) ,
\newblock {\em {2MASS All Sky Catalog of point sources.}}

\bibitem[\protect\astroncite{{Demarque} et~al.}{2004}]{demarque_04}
{Demarque}, P., {Woo}, J.-H., {Kim}, Y.-C., and {Yi}, S.~K. (2004) ,
\newblock {\em \apjs} {\bf 155}, 667

\bibitem[\protect\astroncite{{Dupuy} and {Liu}}{2011}]{dupuy_11}
{Dupuy}, T.~J. and {Liu}, M.~C. (2011) ,
\newblock {\em \apj} {\bf 733}, 122

\bibitem[\protect\astroncite{{Dupuy} et~al.}{2009a}]{dupuy_09}
{Dupuy}, T.~J., {Liu}, M.~C., and {Ireland}, M.~J. (2009a) ,
\newblock {\em \apj} {\bf 692}, 729

\bibitem[\protect\astroncite{{Dupuy} et~al.}{2009b}]{dupuy_09_lhs}
{Dupuy}, T.~J., {Liu}, M.~C., and {Ireland}, M.~J. (2009b) ,
\newblock {\em \apj} {\bf 699}, 168

\bibitem[\protect\astroncite{{Fabregat} and
  {Reglero}}{1990}]{1990AandAS...82..531F}
{Fabregat}, J. and {Reglero}, V. (1990) ,
\newblock {\em \aaps} {\bf 82}, 531

\bibitem[\protect\astroncite{{Fitzgerald} et~al.}{2009}]{fitzgerald_09}
{Fitzgerald}, M.~P., {Kalas}, P.~G., and {Graham}, J.~R. (2009) ,
\newblock {\em \apjl} {\bf 706}, L41

\bibitem[\protect\astroncite{{Ford}}{2006}]{ford_06}
{Ford}, E.~B. (2006) ,
\newblock {\em \apj} {\bf 642}, 505

\bibitem[\protect\astroncite{{Galicher} et~al.}{2011}]{galicher_11}
{Galicher}, R., {Marois}, C., {Macintosh}, B., {Barman}, T., and {Konopacky},
  Q. (2011) ,
\newblock {\em ArXiv e-prints}

\bibitem[\protect\astroncite{{Ghez} et~al.}{2008}]{ghez_08}
{Ghez}, A.~M., {Salim}, S., {Weinberg}, N.~N., {Lu}, J.~R., {Do}, T., {Dunn},
  J.~K., {Matthews}, K., {Morris}, M.~R., {Yelda}, S., {Becklin}, E.~E.,
  {Kremenek}, T., {Milosavljevic}, M., and {Naiman}, J. (2008) ,
\newblock {\em \apj} {\bf 689}, 1044

\bibitem[\protect\astroncite{{Golimowski} et~al.}{2004}]{golimowski_04}
{Golimowski}, D.~A., {Leggett}, S.~K., {Marley}, M.~S., {Fan}, X., {Geballe},
  T.~R., {Knapp}, G.~R., {Vrba}, F.~J., {Henden}, A.~A., {Luginbuhl}, C.~B.,
  {Guetter}, H.~H., {Munn}, J.~A., {Canzian}, B., {Zheng}, W., {Tsvetanov},
  Z.~I., {Chiu}, K., {Glazebrook}, K., {Hoversten}, E.~A., {Schneider}, D.~P.,
  and {Brinkmann}, J. (2004) ,
\newblock {\em \aj} {\bf 127}, 3516

\bibitem[\protect\astroncite{{Hanbury Brown} et~al.}{1974}]{han74}
{Hanbury Brown}, R., {Davis}, J., {Lake}, R.~J.~W., and {Thompson}, R.~J.
  (1974) ,
\newblock {\em \mnras} {\bf 167}, 475

\bibitem[\protect\astroncite{{Hauck} and
  {Mermilliod}}{1998}]{1998AandAS..129..431H}
{Hauck}, B. and {Mermilliod}, M. (1998) ,
\newblock {\em \aaps} {\bf 129}, 431

\bibitem[\protect\astroncite{{Hayward} et~al.}{2001}]{hayward_01}
{Hayward}, T.~L., {Brandl}, B., {Pirger}, B., {Blacken}, C., {Gull}, G.~E.,
  {Schoenwald}, J., and {Houck}, J.~R. (2001) ,
\newblock {\em \pasp} {\bf 113}, 105

\bibitem[\protect\astroncite{{Hinkley} et~al.}{2011}]{hinkley_11_PASP}
{Hinkley}, S., {Oppenheimer}, B.~R., {Zimmerman}, N., {Brenner}, D., {Parry},
  I.~R., {Crepp}, J.~R., {Vasisht}, G., {Ligon}, E., {King}, D., {Soummer}, R.,
  {Sivaramakrishnan}, A., {Beichman}, C., {Shao}, M., {Roberts}, L.~C.,
  {Bouchez}, A., {Dekany}, R., {Pueyo}, L., {Roberts}, J.~E., {Lockhart}, T.,
  {Zhai}, C., {Shelton}, C., and {Burruss}, R. (2011) ,
\newblock {\em \pasp} {\bf 123}, 74

\bibitem[\protect\astroncite{{Howard} et~al.}{2010}]{howard_10}
{Howard}, A.~W., {Marcy}, G.~W., {Johnson}, J.~A., {Fischer}, D.~A., {Wright},
  J.~T., {Isaacson}, H., {Valenti}, J.~A., {Anderson}, J., {Lin}, D.~N.~C., and
  {Ida}, S. (2010) ,
\newblock {\em Science} {\bf 330}, 653

\bibitem[\protect\astroncite{{Ireland} et~al.}{2008}]{ireland_08_BD}
{Ireland}, M.~J., {Kraus}, A., {Martinache}, F., {Lloyd}, J.~P., and {Tuthill},
  P.~G. (2008) ,
\newblock {\em \apj} {\bf 678}, 463

\bibitem[\protect\astroncite{{Isaacson} and
  {Fischer}}{2010}]{isaacson_fischer_10}
{Isaacson}, H. and {Fischer}, D. (2010) ,
\newblock {\em \apj} {\bf 725}, 875

\bibitem[\protect\astroncite{{Janson} et~al.}{2011}]{janson_11}
{Janson}, M., {Carson}, J., {Thalmann}, C., {McElwain}, M.~W., {Goto}, M.,
  {Crepp}, J., {Wisniewski}, J., {Abe}, L., {Brandner}, W., {Burrows}, A.,
  {Egner}, S., {Feldt}, M., {Grady}, C.~A., {Golota}, T., {Guyon}, O.,
  {Hashimoto}, J., {Hayano}, Y., {Hayashi}, M., {Hayashi}, S., {Henning}, T.,
  {Hodapp}, K.~W., {Ishii}, M., {Iye}, M., {Kandori}, R., {Knapp}, G.~R.,
  {Kudo}, T., {Kusakabe}, N., {Kuzuhara}, M., {Matsuo}, T., {Mayama}, S.,
  {Miyama}, S., {Morino}, J., {Moro-Mart{\'{\i}}n}, A., {Nishimura}, T., {Pyo},
  T., {Serabyn}, E., {Suto}, H., {Suzuki}, R., {Takami}, M., {Takato}, N.,
  {Terada}, H., {Tofflemire}, B., {Tomono}, D., {Turner}, E.~L., {Watanabe},
  M., {Yamada}, T., {Takami}, H., {Usuda}, T., and {Tamura}, M. (2011) ,
\newblock {\em \apj} {\bf 728}, 85

\bibitem[\protect\astroncite{{Johnson} and
  {Knuckles}}{1957}]{1957ApJ...126..113J}
{Johnson}, H.~L. and {Knuckles}, C.~F. (1957) ,
\newblock {\em \apj} {\bf 126}, 113

\bibitem[\protect\astroncite{{Johnson} et~al.}{1966}]{1966CoLPL...4...99J}
{Johnson}, H.~L., {Mitchell}, R.~I., {Iriarte}, B., and {Wisniewski}, W.~Z.
  (1966) ,
\newblock {\em Communications of the Lunar and Planetary Laboratory} {\bf 4},
  99

\bibitem[\protect\astroncite{{Johnson} and {Apps}}{2009}]{johnson_09}
{Johnson}, J.~A. and {Apps}, K. (2009) ,
\newblock {\em \apj} {\bf 699}, 933

\bibitem[\protect\astroncite{{Johnson} et~al.}{2011a}]{johnson_11_lhs6343}
{Johnson}, J.~A., {Apps}, K., {Gazak}, J.~Z., {Crepp}, J.~R., {Crossfield},
  I.~J., {Howard}, A.~W., {Marcy}, G.~W., {Morton}, T.~D., {Chubak}, C., and
  {Isaacson}, H. (2011a) ,
\newblock {\em \apj} {\bf 730}, 79

\bibitem[\protect\astroncite{{Johnson} et~al.}{2011b}]{johnson_11_18pack}
{Johnson}, J.~A., {Clanton}, C., {Howard}, A.~W., {Bowler}, B.~P., {Henry},
  G.~W., {Marcy}, G.~W., {Crepp}, J.~R., {Endl}, M., {Cochran}, W.~D.,
  {MacQueen}, P.~J., {Wright}, J.~T., and {Isaacson}, H. (2011b) ,
\newblock {\em ArXiv e-prints}

\bibitem[\protect\astroncite{{Jones} et~al.}{1994}]{jones_94}
{Jones}, H.~R.~A., {Longmore}, A.~J., {Jameson}, R.~F., and {Mountain}, C.~M.
  (1994) ,
\newblock {\em \mnras} {\bf 267}, 413

\bibitem[\protect\astroncite{{Kirkpatrick} et~al.}{1999a}]{kirkpatrick_99}
{Kirkpatrick}, J.~D., {Allard}, F., {Bida}, T., {Zuckerman}, B., {Becklin},
  E.~E., {Chabrier}, G., and {Baraffe}, I. (1999a) ,
\newblock {\em \apj} {\bf 519}, 834

\bibitem[\protect\astroncite{{Kirkpatrick} et~al.}{1999b}]{kirkpatrick_99_lt}
{Kirkpatrick}, J.~D., {Reid}, I.~N., {Liebert}, J., {Cutri}, R.~M., {Nelson},
  B., {Beichman}, C.~A., {Dahn}, C.~C., {Monet}, D.~G., {Gizis}, J.~E., and
  {Skrutskie}, M.~F. (1999b) ,
\newblock {\em \apj} {\bf 519}, 802

\bibitem[\protect\astroncite{{Kirkpatrick} et~al.}{2000}]{kirkpatrick_00}
{Kirkpatrick}, J.~D., {Reid}, I.~N., {Liebert}, J., {Gizis}, J.~E.,
  {Burgasser}, A.~J., {Monet}, D.~G., {Dahn}, C.~C., {Nelson}, B., and
  {Williams}, R.~J. (2000) ,
\newblock {\em \aj} {\bf 120}, 447

\bibitem[\protect\astroncite{{Konopacky} et~al.}{2010}]{konopacky_10}
{Konopacky}, Q.~M., {Ghez}, A.~M., {Barman}, T.~S., {Rice}, E.~L., {Bailey},
  III, J.~I., {White}, R.~J., {McLean}, I.~S., and {Duch{\^e}ne}, G. (2010) ,
\newblock {\em \apj} {\bf 711}, 1087

\bibitem[\protect\astroncite{{Lafreni{\`e}re} et~al.}{2007}]{lafreniere_loci}
{Lafreni{\`e}re}, D., {Marois}, C., {Doyon}, R., {Nadeau}, D., and {Artigau},
  {\'E}. (2007) ,
\newblock {\em \apj} {\bf 660}, 770

\bibitem[\protect\astroncite{{Lagrange} et~al.}{2010}]{lagrange_10}
{Lagrange}, A., {Bonnefoy}, M., {Chauvin}, G., {Apai}, D., {Ehrenreich}, D.,
  {Boccaletti}, A., {Gratadour}, D., {Rouan}, D., {Mouillet}, D., {Lacour}, S.,
  and {Kasper}, M. (2010) ,
\newblock {\em ArXiv e-prints}

\bibitem[\protect\astroncite{{Lane} et~al.}{2001}]{lane_01}
{Lane}, B.~F., {Zapatero Osorio}, M.~R., {Britton}, M.~C., {Mart{\'{\i}}n},
  E.~L., and {Kulkarni}, S.~R. (2001) ,
\newblock {\em \apj} {\bf 560}, 390

\bibitem[\protect\astroncite{{Liu} et~al.}{2011}]{liu_11}
{Liu}, M.~C., {Delorme}, P., {Dupuy}, T.~J., {Bowler}, B.~P., {Albert}, L.,
  {Artigau}, E., {Reyle}, C., {Forveille}, T., and {Delfosse}, X. (2011) ,
\newblock {\em ArXiv e-prints}

\bibitem[\protect\astroncite{{Liu} et~al.}{2002}]{liu_02}
{Liu}, M.~C., {Fischer}, D.~A., {Graham}, J.~R., {Lloyd}, J.~P., {Marcy},
  G.~W., and {Butler}, R.~P. (2002) ,
\newblock {\em \apj} {\bf 571}, 519

\bibitem[\protect\astroncite{{Liu} et~al.}{2007}]{liu_07}
{Liu}, M.~C., {Leggett}, S.~K., and {Chiu}, K. (2007) ,
\newblock {\em \apj} {\bf 660}, 1507

\bibitem[\protect\astroncite{{Lu} et~al.}{2009}]{lu_09}
{Lu}, J.~R., {Ghez}, A.~M., {Hornstein}, S.~D., {Morris}, M.~R., {Becklin},
  E.~E., and {Matthews}, K. (2009) ,
\newblock {\em \apj} {\bf 690}, 1463

\bibitem[\protect\astroncite{{Luhman} et~al.}{2011}]{luhman_11}
{Luhman}, K.~L., {Burgasser}, A.~J., and {Bochanski}, J.~J. (2011) ,
\newblock {\em \apjl} {\bf 730}, L9

\bibitem[\protect\astroncite{{Marcy} and {Butler}}{1992}]{marcy_butler_92}
{Marcy}, G.~W. and {Butler}, R.~P. (1992) ,
\newblock {\em \pasp} {\bf 104}, 270

\bibitem[\protect\astroncite{{Marley} et~al.}{2010}]{marley_10}
{Marley}, M.~S., {Saumon}, D., and {Goldblatt}, C. (2010) ,
\newblock {\em \apjl} {\bf 723}, L117

\bibitem[\protect\astroncite{{Marley} et~al.}{2002}]{marley_02}
{Marley}, M.~S., {Seager}, S., {Saumon}, D., {Lodders}, K., {Ackerman}, A.~S.,
  {Freedman}, R.~S., and {Fan}, X. (2002) ,
\newblock {\em \apj} {\bf 568}, 335

\bibitem[\protect\astroncite{{Marois} et~al.}{2006}]{marois_06}
{Marois}, C., {Lafreni{\`e}re}, D., {Doyon}, R., {Macintosh}, B., and {Nadeau},
  D. (2006) ,
\newblock {\em \apj} {\bf 641}, 556

\bibitem[\protect\astroncite{{McClure} and
  {Forrester}}{1981}]{1981PDAO...15..439M}
{McClure}, R.~D. and {Forrester}, W.~T. (1981) ,
\newblock {\em Publications of the Dominion Astrophysical Observatory Victoria}
  {\bf 15}, 439

\bibitem[\protect\astroncite{{Mermilliod}}{1986}]{1986EgUBV........0M}
{Mermilliod}, J.-C. (1986) ,
\newblock {\em Catalogue of Eggen's UBV data., 0 (1986)} pp. 0--+

\bibitem[\protect\astroncite{{Metchev} and
  {Hillenbrand}}{2006}]{metchev_hillenbrand_06}
{Metchev}, S.~A. and {Hillenbrand}, L.~A. (2006) ,
\newblock {\em \apj} {\bf 651}, 1166

\bibitem[\protect\astroncite{{Moffett} and
  {Barnes}}{1980}]{1980ApJS...44..427M}
{Moffett}, T.~J. and {Barnes}, III, T.~G. (1980) ,
\newblock {\em \apjs} {\bf 44}, 427

\bibitem[\protect\astroncite{{Nakajima} et~al.}{1995}]{nakajima_95}
{Nakajima}, T., {Oppenheimer}, B.~R., {Kulkarni}, S.~R., {Golimowski}, D.~A.,
  {Matthews}, K., and {Durrance}, S.~T. (1995) ,
\newblock {\em \nat} {\bf 378}, 463

\bibitem[\protect\astroncite{{Nicolet}}{1978}]{1978AandAS...34....1N}
{Nicolet}, B. (1978) ,
\newblock {\em \aaps} {\bf 34}, 1

\bibitem[\protect\astroncite{{Niconov} et~al.}{1957}]{1957IzKry..17...42N}
{Niconov}, V.~B., {Nekrasova}, S.~V., {Polosuina}, N.~S., {Rachkouvsky}, N.~D.,
  and {Chuvajev}, W.~K. (1957) ,
\newblock {\em Izvestiya Ordena Trudovogo Krasnogo Znameni Krymskoj
  Astrofizicheskoj Observatorii} {\bf 17}, 42

\bibitem[\protect\astroncite{{Noyes} et~al.}{1984}]{noyes_84}
{Noyes}, R.~W., {Hartmann}, L.~W., {Baliunas}, S.~L., {Duncan}, D.~K., and
  {Vaughan}, A.~H. (1984) ,
\newblock {\em \apj} {\bf 279}, 763

\bibitem[\protect\astroncite{{Olsen}}{1993}]{1993AandAS..102...89O}
{Olsen}, E.~H. (1993) ,
\newblock {\em \aaps} {\bf 102}, 89

\bibitem[\protect\astroncite{{Olsen}}{1994}]{1994AandAS..106..257O}
{Olsen}, E.~H. (1994) ,
\newblock {\em \aaps} {\bf 106}, 257

\bibitem[\protect\astroncite{{Oppenheimer} and
  {Hinkley}}{2009}]{oppenheimer_hinkley_09}
{Oppenheimer}, B.~R. and {Hinkley}, S. (2009) ,
\newblock {\em \araa} {\bf 47}, 253

\bibitem[\protect\astroncite{{Oppenheimer} et~al.}{1995}]{oppenheimer_95}
{Oppenheimer}, B.~R., {Kulkarni}, S.~R., {Matthews}, K., and {Nakajima}, T.
  (1995) ,
\newblock {\em Science} {\bf 270}, 1478

\bibitem[\protect\astroncite{{Pickles}}{1998}]{pic98}
{Pickles}, A.~J. (1998) ,
\newblock {\em \pasp} {\bf 110}, 863

\bibitem[\protect\astroncite{{Pinfield} et~al.}{2006}]{pinfield_06}
{Pinfield}, D.~J., {Jones}, H.~R.~A., {Lucas}, P.~W., {Kendall}, T.~R.,
  {Folkes}, S.~L., {Day-Jones}, A.~C., {Chappelle}, R.~J., and {Steele}, I.~A.
  (2006) ,
\newblock {\em \mnras} {\bf 368}, 1281

\bibitem[\protect\astroncite{{Rebolo} et~al.}{1995}]{rebolo_95}
{Rebolo}, R., {Zapatero Osorio}, M.~R., and {Mart{\'{\i}}n}, E.~L. (1995) ,
\newblock {\em \nat} {\bf 377}, 129

\bibitem[\protect\astroncite{{Rudy} et~al.}{1996}]{rudy_96}
{Rudy}, R.~J., {Rossano}, G.~S., and {Puetter}, R.~C. (1996) ,
\newblock {\em \apjl} {\bf 458}, L41

\bibitem[\protect\astroncite{{Rufener}}{1976}]{1976AandAS...26..275R}
{Rufener}, F. (1976) ,
\newblock {\em \aaps} {\bf 26}, 275

\bibitem[\protect\astroncite{{Sahlmann} et~al.}{2011}]{sahlmann_11}
{Sahlmann}, J., {Lovis}, C., {Queloz}, D., and {S{\'e}gransan}, D. (2011) ,
\newblock {\em \aap} {\bf 528}, L8+

\bibitem[\protect\astroncite{{Saumon} and {Marley}}{2008}]{saumon_08}
{Saumon}, D. and {Marley}, M.~S. (2008) ,
\newblock {\em \apj} {\bf 689}, 1327

\bibitem[\protect\astroncite{{Serabyn} et~al.}{2009}]{serabyn_09}
{Serabyn}, E., {Mawet}, D., {Bloemhof}, E., {Haguenauer}, P., {Mennesson}, B.,
  {Wallace}, K., and {Hickey}, J. (2009) ,
\newblock {\em \apj} {\bf 696}, 40

\bibitem[\protect\astroncite{{Stassun} et~al.}{2006}]{stassun_06}
{Stassun}, K.~G., {Mathieu}, R.~D., and {Valenti}, J.~A. (2006) ,
\newblock {\em \nat} {\bf 440}, 311

\bibitem[\protect\astroncite{{Stevenson}}{1991}]{stevenson_91}
{Stevenson}, D.~J. (1991) ,
\newblock {\em \araa} {\bf 29}, 163

\bibitem[\protect\astroncite{{ten Brummelaar} et~al.}{2005}]{ten05}
{ten Brummelaar}, T.~A., {McAlister}, H.~A., {Ridgway}, S.~T., {Bagnuolo}, Jr.,
  W.~G., {Turner}, N.~H., {Sturmann}, L., {Sturmann}, J., {Berger}, D.~H.,
  {Ogden}, C.~E., {Cadman}, R., {Hartkopf}, W.~I., {Hopper}, C.~H., and
  {Shure}, M.~A. (2005) ,
\newblock {\em \apj} {\bf 628}, 453

\bibitem[\protect\astroncite{{Torres} et~al.}{2010}]{torres_10}
{Torres}, G., {Andersen}, J., and {Gim{\'e}nez}, A. (2010) ,
\newblock {\em \aapr} {\bf 18}, 67

\bibitem[\protect\astroncite{{Valenti} et~al.}{2009}]{valenti_09}
{Valenti}, J.~A., {Fischer}, D., {Marcy}, G.~W., {Johnson}, J.~A., {Henry},
  G.~W., {Wright}, J.~T., {Howard}, A.~W., {Giguere}, M., and {Isaacson}, H.
  (2009) ,
\newblock {\em \apj} {\bf 702}, 989

\bibitem[\protect\astroncite{{Valenti} and
  {Fischer}}{2005}]{valenti_fischer_05}
{Valenti}, J.~A. and {Fischer}, D.~A. (2005) ,
\newblock {\em \apjs} {\bf 159}, 141

\bibitem[\protect\astroncite{{van Belle} et~al.}{2007}]{vcb07}
{van Belle}, G.~T., {Ciardi}, D.~R., and {Boden}, A.~F. (2007) ,
\newblock {\em \apj} {\bf 657}, 1058

\bibitem[\protect\astroncite{{van Leeuwen}}{2007}]{van_leeuwen_07}
{van Leeuwen}, F. (2007) ,
\newblock {\em \aap} {\bf 474}, 653

\bibitem[\protect\astroncite{{Vogt} et~al.}{1994}]{vogt_94}
{Vogt}, S.~S., {Allen}, S.~L., {Bigelow}, B.~C., {Bresee}, L., {Brown}, B.,
  {Cantrall}, T., {Conrad}, A., {Couture}, M., {Delaney}, C., {Epps}, H.~W.,
  {Hilyard}, D., {Hilyard}, D.~F., {Horn}, E., {Jern}, N., {Kanto}, D.,
  {Keane}, M.~J., {Kibrick}, R.~I., {Lewis}, J.~W., {Osborne}, J.,
  {Pardeilhan}, G.~H., {Pfister}, T., {Ricketts}, T., {Robinson}, L.~B.,
  {Stover}, R.~J., {Tucker}, D., {Ward}, J., and {Wei}, M.~Z. (1994) ,
\newblock In {\em Society of Photo-Optical Instrumentation Engineers (SPIE)
  Conference Series}.  ({D.~L.~Crawford \& E.~R.~Craine} ed.), Vol. 2198 of
  {\em Society of Photo-Optical Instrumentation Engineers (SPIE) Conference
  Series}, pp. 362--+

\bibitem[\protect\astroncite{{von Braun} et~al.}{2011a}]{von11}
{von Braun}, K., {Boyajian}, T.~S., {Kane}, S.~R., {van Belle}, G.~T.,
  {Ciardi}, D.~R., {L{\'o}pez-Morales}, M., {McAlister}, H.~A., {Henry}, T.~J.,
  {Jao}, W., {Riedel}, A.~R., {Subasavage}, J.~P., {Schaefer}, G., {ten
  Brummelaar}, T.~A., {Ridgway}, S., {Sturmann}, L., {Sturmann}, J.,
  {Mazingue}, J., {Turner}, N.~H., {Farrington}, C., {Goldfinger}, P.~J., and
  {Boden}, A.~F. (2011a) ,
\newblock {\em \apjl} {\bf 729}, L26+

\bibitem[\protect\astroncite{{von Braun} et~al.}{2011b}]{von11c}
{von Braun}, K., {Boyajian}, T.~S., {ten Brummelaar}, T.~A., {Kane}, S.~R.,
  {van Belle}, G.~T., {Ciardi}, D.~R., {Raymond}, S.~N., {Lopez-Morales}, M.,
  {McAlister}, H.~A., {Schaefer}, G., {Ridgway}, S.~T., {Sturmann}, L.,
  {Sturmann}, J., {White}, R., {Turner}, N.~H., {Farrington}, C., and
  {Goldfinger}, P.~J. (2011b) ,
\newblock {\em ArXiv e-prints: astro-ph/1106.1152}

\bibitem[\protect\astroncite{{Wahhaj} et~al.}{2011}]{wahhaj_11}
{Wahhaj}, Z., {Liu}, M.~C., {Biller}, B.~A., {Clarke}, F., {Nielsen}, E.~L.,
  {Close}, L.~M., {Hayward}, T.~L., {Mamajek}, E.~E., {Cushing}, M., {Dupuy},
  T., {Tecza}, M., {Thatte}, N., {Chun}, M., {Ftaclas}, C., {Hartung}, M.,
  {Reid}, I.~N., {Shkolnik}, E.~L., {Alencar}, S.~H.~P., {Artymowicz}, P.,
  {Boss}, A., {de Gouveia Dal Pino}, E., {Gregorio-Hetem}, J., {Ida}, S.,
  {Kuchner}, M., {Lin}, D.~N.~C., and {Toomey}, D.~W. (2011) ,
\newblock {\em ArXiv e-prints}

\bibitem[\protect\astroncite{{Wright} et~al.}{2004}]{wright_04}
{Wright}, J.~T., {Marcy}, G.~W., {Butler}, R.~P., and {Vogt}, S.~S. (2004) ,
\newblock {\em \apjs} {\bf 152}, 261

\bibitem[\protect\astroncite{{Zapatero Osorio} et~al.}{2004}]{zapatero_04}
{Zapatero Osorio}, M.~R., {Lane}, B.~F., {Pavlenko}, Y., {Mart{\'{\i}}n},
  E.~L., {Britton}, M., and {Kulkarni}, S.~R. (2004) ,
\newblock {\em \apj} {\bf 615}, 958

\end{thebibliography}
\end{small}

\begin{table*}[!t]
\centerline{
\begin{tabular}{rccc}
\hline
\hline
Dewar             &    Telescope     &     RV offset (m/s)              &   RV jitter (m/s) \\
\hline
      6                &     Lick              &      --                                  &    $24.3^{+4.5}_{-5.4}$           \\
      8                &     Lick              &    $28.4^{+5.5}_{-6.9}$      &     $9.8^{+1.9}_{-2.7}$            \\               
    18                &    Lick               &    $38.8^{+7.2}_{-8.6}$      &     $8.5^{+1.4}_{-1.8}$            \\
    24                &    Lick               &    $63.5^{+9.4}_{-11.4}$     &    $9.2^{+2.1}_{-2.7}$            \\
    39                &     Lick              &    $46.9^{+6.0}_{-7.3}$       &   $9.3^{+1.8}_{-2.3}$              \\  
  102                &    Keck             &     $50.7^{+6.8}_{-8.1}$       &    $7.1^{+0.9}_{-1.1}$             \\
  103                &    Keck             &     $65.0^{+10.0}_{-11.9}$   &    $5.6^{+0.7}_{-0.9}$           \\
\hline
\hline
\end{tabular}}
\caption{RV offsets and jitter values derived from our MCMC analysis (68.2\% confidence interval).} 
\label{tab:offsets}
\end{table*}

\begin{deluxetable}{rcccc}
\tablecaption{Doppler Radial Velocity Data \label{tab:rvdata}} 
\tablewidth{0pc}
\tablehead{\colhead{JD-2,440,000} & \colhead{RV (m/s)} & \colhead{error (m/s)}  & \colhead{Instrument}  &  \colhead{Dewar}	
}
\startdata
7047.7194		&	270.71		&	0.81	&	L	&	6	\\
7373.8210		&	232.88		&	0.38	&	L	&	6	\\
7373.9403		&	222.46		&	0.68	&	L	&	6	\\
7374.8270		&	208.96		&	0.63	&	L	&	6	\\
7374.9631		&	241.49		&	0.87	&	L	&	6	\\
7394.7872		&	223.48		&	1.29	&	L	&	6	\\
7430.6837		&	260.22		&	1.02	&	L	&	6	\\
7431.6884		&	273.80		&	1.10	&	L	&	6	\\
7578.0606		&	247.64		&	1.29	&	L	&	6	\\
7710.8762		&	232.07		&	1.29	&	L	&	6	\\
7846.7068		&	201.73		&	1.03	&	L	&	6	\\
8018.9895		&	191.07		&	1.64	&	L	&	6	\\
8019.9631		&	216.91		&	1.63	&	L	&	6	\\
8113.8532		&	150.19		&	1.02	&	L	&	6	\\
8200.6134		&	176.47		&	1.03	&	L	&	6	\\
8375.0093		&	158.68		&	1.63	&	L	&	6	\\
8437.9603		&	173.20		&	1.53	&	L	&	6	\\
8834.8294		&	133.43	      &	0.75	&	L	&	8	\\
8847.8253		&	128.49	      &	0.93	&	L	&	8	\\
8905.7100		&	136.85	      &	0.84	&	L	&	8	\\
9123.0038		&	91.09		&	1.38	&	L	&	8	\\
9171.8390		&	103.59	      &	1.11	&	L	&	8	\\
9173.8394		&	98.98		&	1.26	&	L	&	8	\\
9174.9026		&	110.75	      &	1.12	&	L	&	8	\\
9200.8275		&	100.12	      &	1.04	&	L	&	8	\\
9278.7378		&	105.00	      &	1.08	&	L	&	8	\\
9280.7483		&	97.92		&	0.99	&	L	&	8	\\
9587.7636		&	80.56		&	0.98	&	L	&	8	\\
9589.7614		&	82.45		&	1.03	&	L	&	8	\\
9602.7461		&	101.85	      &	0.96	&	L	&	8	\\
9622.7247		&	90.93		&	1.26	&	L	&	8	\\
9680.5946		&	66.84		&	0.76	&	L	&	39	\\
9872.9322		&	54.42		&	0.49	&	L	&	39	\\
9893.8756		&	61.43		&	0.50	&	L	&	39	\\
9913.9177		&	45.84		&	0.51	&	L	&	39	\\
9984.6846		&	56.55		&	0.51	&	L	&	39	\\
10215.8672	&	8.85		&	0.78	&	L	&	39	\\
10263.8492	&	26.16		&	0.54	&	L	&	39	\\
10299.8286	&	33.72		&	0.61	&	L	&	39	\\
10304.8468	&	18.14		&	0.56	&	L	&	39	\\
10601.8849	&	0.00		&	0.01	&	K	&	102	\\
10602.9911	&	3.78		&	1.29	&	K	&	102	\\
10604.0217	&	3.69		&	1.19	&	K	&	102	\\
10605.0298	&	2.92		&	1.30	&	K	&	102	\\
10606.0621	&	-5.73		&	1.30	&	K	&	102	\\
10606.9658	&	0.77		&	1.38	&	K	&	102	\\
10607.9157	&	3.79		&	1.37	&	K	&	102	\\
10613.9009	&	-2.26		&	0.79	&	L	&	39	\\
10614.9313	&	-0.37		&	0.75	&	L	&	39	\\
10640.8998	&	6.26		&	0.75	&	L	&	39	\\
10656.8269	&	7.41		&	0.76	&	L	&	39	\\
10793.5628	&	1.76		&	0.51	&	L	&	39	\\
10979.9278	&	-6.21		&	1.16	&	L	&	18	\\
11048.8474	&	5.92		&	1.17	&	L	&	18	\\
11062.7690	&	-9.13		&	1.23	&	L	&	18	\\
11068.8450	&	-23.71	&	1.45	&	K	&	102	\\
11069.8909	&	-29.20	&	1.34	&	K	&	102	\\
11070.9108	&	-19.72	&	1.31	&	K	&	102	\\
11071.8471	&	-13.15	&	1.60	&	K	&	102	\\
11072.8352	&	-15.16	&	1.34	&	K	&	102	\\
11074.8549	&	-22.96	&	1.31	&	K	&	102	\\
11075.7848	&	-22.30	&	1.18	&	K	&	102	\\
11300.0046	&	-33.20	&	0.78	&	L	&	39	\\
11303.9778	&	-33.21	&	0.80	&	L	&	18	\\
11305.9730	&	-32.72	&	0.88	&	L	&	18	\\
11311.1002	&	-41.88	&	1.23	&	K	&	102	\\
11312.0886	&	-40.53	&	1.37	&	K	&	102	\\
11313.1019	&	-33.60	&	1.39	&	K	&	102	\\
11314.1248	&	-39.07	&	1.04	&	K	&	102	\\
11367.9059	&	-53.11	&	1.51	&	K	&	102	\\
11368.8710	&	-49.16	&	1.72	&	K	&	102	\\
11370.0152	&	-45.70	&	1.28	&	K	&	102	\\
11372.0268	&	-41.96	&	1.64	&	K	&	102	\\
11373.0623	&	-46.15	&	1.59	&	K	&	102	\\
11373.8199	&	-51.10	&	1.48	&	K	&	102	\\
11389.8655	&	-51.27	&	0.80	&	L	&	18	\\
11392.8306	&	-42.50	&	0.89	&	L	&	18	\\
11410.8762	&	-52.99	&	1.52	&	K	&	102	\\
11411.8717	&	-55.86	&	1.39	&	K	&	102	\\
11416.7036	&	-29.17	&	0.88	&	L	&	18	\\
11439.8221	&	-53.96	&	1.46	&	K	&	102	\\
11467.6397	&	-28.58	&	1.57	&	L	&	18	\\
11468.6072	&	-34.72	&	1.31	&	L	&	18	\\
11679.0915	&	-39.86	&	1.85	&	K	&	102	\\
11703.0748	&	-46.32	&	1.50	&	K	&	102	\\
11751.8211	&	-50.03	&	0.88	&	L	&	18	\\
11754.9418	&	-49.21	&	1.33	&	K	&	102	\\
11793.8250	&	-53.33	&	1.63	&	K	&	102	\\
11815.6476	&	-45.26	&	1.48	&	L	&	18	\\
12031.0398	&	-79.77	&	1.53	&	K	&	102	\\
12041.9946	&	-78.96	&	1.22	&	L	&	18	\\
12075.9299	&	-81.23	&	0.98	&	L	&	18	\\
12099.0483	&	-87.46	&	1.60	&	K	&	102	\\
12115.8858	&	-87.66	&	0.95	&	L	&	18	\\
12117.8641	&	-82.71	&	0.82	&	L	&	18	\\
12127.9831	&	-100.94    &	1.83	&	K	&	102	\\
12133.8002	&	-98.58	    &	1.62	&	K	&	102	\\
12140.7744	&	-69.50	    &	0.87	&	L	&	18	\\
12181.6483	&	-84.26	    &	0.99	&	L	&	18	\\
12186.6687	&	-75.66	    &	1.15	&	L	&	18	\\
12188.7009	&	-77.17	    &	1.29	&	L	&	18	\\
12202.6766	&	-91.70	    &	1.09	&	L	&	18	\\
12391.1392	&	-120.49	&	1.91	&	K	&	102	\\
12508.7589	&	-120.32	&	1.18	&	L	&	24	\\
12509.7462	&	-134.04	&	1.14	&	L	&	24	\\
12515.7825	&	-115.41	&	1.12	&	K	&	102	\\
12575.6912	&	-116.79	&	1.69	&	K	&	102	\\
12848.9031	&	-131.74	&	1.78	&	K	&	102	\\
12855.9988	&	-136.42	&	1.67	&	K	&	102	\\
12894.7297	&	-139.06	&	1.45	&	L	&	24	\\
13181.0272	&	-171.06	&	1.48	&	K	&	102	\\
13203.8281	&	-160.71	&	1.90	&	L	&	24	\\
13303.7113	&	-172.17	&	0.84	&	K	&	103	\\
13544.9448	&	-203.04	&	1.47	&	L	&	24	\\
13551.0081	&	-195.05	&	0.55	&	K	&	103	\\
13556.9127	&	-190.72	&	1.37	&	L	&	24	\\
13561.8937	&	-201.13	&	1.58	&	L	&	24	\\
13589.8255	&	-186.90	&	2.81	&	L	&	24	\\
13603.9205	&	-195.75	&	0.74	&	K	&	103	\\
13926.0386	&	-212.10	&	0.67	&	K	&	103	\\
13926.8957	&	-217.61	&	1.65	&	L	&	24	\\
13954.7593	&	-226.25	&	1.97	&	L	&	24	\\
13958.7828	&	-221.32	&	1.29	&	L	&	24	\\
13982.8138	&	-220.35	&	0.62	&	K	&	103	\\
13989.8700	&	-224.79	&	3.04	&	L	&	24	\\
14337.0434	&	-235.41	&	1.44	&	K	&	103	\\
14546.1518	&	-240.05	&	1.22	&	K	&	103	\\
14671.9587	&	-240.90	&	1.21	&	K	&	103	\\
14673.9656	&	-248.17	&	1.16	&	K	&	103	\\
14675.8841	&	-247.46	&	0.93	&	K	&	103	\\
14688.8742	&	-239.27	&	1.47	&	K	&	103	\\
14689.9519	&	-241.90	&	1.36	&	K	&	103	\\
14697.9197	&	-222.74	&	2.94	&	L	&	24	\\
14717.9307	&	-244.81	&	1.25	&	K	&	103	\\
14718.9864	&	-246.41	&	1.27	&	K	&	103	\\
14719.9911	&	-239.01	&	1.18	&	K	&	103	\\
14720.9566	&	-225.02	&	1.22	&	K	&	103	\\
14721.9701	&	-218.99	&	1.19	&	K	&	103	\\
14722.8696	&	-226.22	&	1.20	&	K	&	103	\\
14724.9263	&	-233.32	&	1.27	&	K	&	103	\\
14725.8416	&	-232.83	&	1.23	&	K	&	103	\\
14726.9581	&	-231.17	&	1.16	&	K	&	103	\\
14727.8365	&	-236.50	&	1.21	&	K	&	103	\\
14777.8159	&	-235.80	&	1.43	&	K	&	103	\\
14808.7068	&	-234.60	&	1.35	&	K	&	103	\\
14929.1287	&	-234.40	&	1.43	&	K	&	103	\\
14930.1221	&	-237.22	&	1.34	&	K	&	103	\\
14935.1295	&	-230.98	&	1.22	&	K	&	103	\\
15041.8701	&	-229.55	&	1.34	&	K	&	103	\\
15076.7430	&	-230.53	&	1.29	&	K	&	103	\\
15079.7372	&	-232.51	&	1.34	&	K	&	103	\\
15106.9076	&	-219.21	&	1.36	&	K	&	103	\\
15169.7567	&	-230.30	&	0.88	&	K	&	103	\\
15290.1483	&	-218.21	&	1.36	&	K	&	103	\\
15313.1369	&	-224.11	&	1.32	&	K	&	103	\\
15319.0997	&	-226.52	&	1.48	&	K	&	103	\\
15404.9095	&	-214.63	&	1.35	&	K	&	103	\\
15437.0096	&	-206.43	&	1.35	&	K	&	103	\\
15542.6875	&	-198.94	&	1.15	&	K	&	103	\\
15723.0122	&	-163.58	&	1.38	&	K	&	103	\\
15785.9666	&	-169.26	&	1.58	&	K	&	103	\\
15842.7963     &   -158.36      &   1.57     &    K    &   103     \\
\enddata
\tablecomments{Raw RV measurements used for orbital analysis. Uncertainties correspond to the error due to photon noise only. Observations obtained at Lick and Keck are labeled in the instrument column. Several different hardware configurations have been used over the past 24 years. They are labeled according to dewar number. Each require different RV offset and jitter values.}
\end{deluxetable}

\begin{deluxetable}{rcc}
\tablecaption{Literature Photometry for HR7672 \label{tab:photometry}} 
\tablewidth{0pc}
\tablehead{\colhead{Filter} & \colhead{Value (mag)} & \colhead{Reference}	
}
\startdata
 Johnson $V$ & 5.80 $\pm$ 0.05 & \citet{1966CoLPL...4...99J} \\
 Johnson $B$ & 6.410 $\pm$ 0.05 & \citet{1966CoLPL...4...99J} \\
 Johnson $U$ & 6.500 $\pm$ 0.05 & \citet{1966CoLPL...4...99J} \\
 Johnson $V$ & 5.80 $\pm$ 0.05 & \citet{1967AJ.....72.1334C} \\
 Johnson $B$ & 6.390 $\pm$ 0.05 & \citet{1967AJ.....72.1334C} \\
 Johnson $U$ & 6.500 $\pm$ 0.05 & \citet{1967AJ.....72.1334C} \\
 Johnson $V$ & 5.80 $\pm$ 0.05 & \citet{ 1957ApJ...126..113J} \\
 Johnson $B$ & 6.410 $\pm$ 0.05 & \citet{1957ApJ...126..113J} \\
 Johnson $U$ & 6.500 $\pm$ 0.05 & \citet{1957ApJ...126..113J} \\
 Johnson $V$ & 5.77 $\pm$ 0.05 & \citet{1957IzKry..17...42N} \\
 Johnson $B$ & 6.360 $\pm$ 0.05 & \citet{1957IzKry..17...42N} \\
 Johnson $V$ & 5.79 $\pm$ 0.05 & \citet{1986EgUBV........0M} \\
 Johnson $B$ & 6.390 $\pm$ 0.05 & \citet{1986EgUBV........0M} \\
 Johnson $U$ & 6.490 $\pm$ 0.05 & \citet{1986EgUBV........0M} \\
 Johnson $V$ & 5.799 $\pm$ 0.05 & \citet{1980ApJS...44..427M} \\
 Johnson $B$ & 6.402 $\pm$ 0.05 & \citet{1980ApJS...44..427M} \\
 Johnson $V$ & 5.80 $\pm$ 0.05 & \citet{1978AandAS...34....1N} \\
 Johnson $B$ & 6.410 $\pm$ 0.05 & \citet{1978AandAS...34....1N} \\
 Johnson $U$ & 6.500 $\pm$ 0.05 & \citet{1978AandAS...34....1N} \\
 Johnson $U$ & 6.5 $\pm$ 0.05 & \citet{1997PASP..109..645B} \\
 Johnson $U$ & 6.5 $\pm$ 0.05 & \citet{1997PASP..109..645B} \\
 Johnson $U$ & 6.53 $\pm$ 0.05 & \citet{1997PASP..109..645B} \\
 Johnson $U$ & 6.5 $\pm$ 0.05 & \citet{1997PASP..109..645B} \\
 Johnson $U$ & 6.49 $\pm$ 0.05 & \citet{1997PASP..109..645B} \\
 Johnson $B$ & 6.41 $\pm$ 0.05 & \citet{1997PASP..109..645B} \\
 Johnson $B$ & 6.39 $\pm$ 0.05 & \citet{1997PASP..109..645B} \\
 Johnson $B$ & 6.41 $\pm$ 0.05 & \citet{1997PASP..109..645B} \\
 Johnson $B$ & 6.41 $\pm$ 0.05 & \citet{1997PASP..109..645B} \\
 Johnson $B$ & 6.36 $\pm$ 0.05 & \citet{1997PASP..109..645B} \\
 Johnson $B$ & 6.394 $\pm$ 0.05 & \citet{1997PASP..109..645B} \\
 Johnson $B$ & 6.39 $\pm$ 0.05 & \citet{1997PASP..109..645B} \\
 Johnson $B$ & 6.402 $\pm$ 0.05 & \citet{1997PASP..109..645B} \\
 Johnson $V$ & 5.8 $\pm$ 0.05 & \citet{1997PASP..109..645B} \\
 Johnson $V$ & 5.8 $\pm$ 0.05 & \citet{1997PASP..109..645B} \\
 Johnson $V$ & 5.79 $\pm$ 0.05 & \citet{1997PASP..109..645B} \\
 Johnson $V$ & 5.8 $\pm$ 0.05 & \citet{1997PASP..109..645B} \\
 Johnson $V$ & 5.77 $\pm$ 0.05 & \citet{1997PASP..109..645B} \\
 Johnson $V$ & 5.797 $\pm$ 0.05 & \citet{1997PASP..109..645B} \\
 Johnson $V$ & 5.79 $\pm$ 0.05 & \citet{1997PASP..109..645B} \\
 Johnson $V$ & 5.799 $\pm$ 0.05 & \citet{1997PASP..109..645B} \\
 Stromgren $u$ & 7.681 $\pm$ 0.08 & \citet{1998AandAS..129..431H} \\
 Stromgren $v$ & 6.765 $\pm$ 0.08 & \citet{1998AandAS..129..431H} \\
 Stromgren $b$ & 6.187 $\pm$ 0.08 & \citet{1998AandAS..129..431H} \\
 Stromgren $y$ & 5.80 $\pm$ 0.08 & \citet{1998AandAS..129..431H} \\
 Stromgren $u$ & 7.673 $\pm$ 0.08 & \citet{1994AandAS..106..257O} \\
 Stromgren $v$ & 6.757 $\pm$ 0.08 & \citet{1994AandAS..106..257O} \\
 Stromgren $b$ & 6.184 $\pm$ 0.08 & \citet{1994AandAS..106..257O} \\
 Stromgren $y$ & 5.80 $\pm$ 0.08 & \citet{1994AandAS..106..257O} \\
 Stromgren $u$ & 7.675 $\pm$ 0.08 & \citet{1990AandAS...82..531F} \\
 Stromgren $v$ & 6.761 $\pm$ 0.08 & \citet{1990AandAS...82..531F} \\
 Stromgren $b$ & 6.184 $\pm$ 0.08 & \citet{1990AandAS...82..531F} \\
 Stromgren $y$ & 5.80 $\pm$ 0.08 & \citet{1990AandAS...82..531F} \\
 Stromgren $u$ & 7.682 $\pm$ 0.08 & \citet{1993AandAS..102...89O} \\
 Stromgren $v$ & 6.761 $\pm$ 0.08 & \citet{1993AandAS..102...89O} \\
 Stromgren $b$ & 6.187 $\pm$ 0.08 & \citet{1993AandAS..102...89O} \\
 Stromgren $y$ & 5.80 $\pm$ 0.08 & \citet{1993AandAS..102...89O} \\
 Stromgren $u$ & 7.682 $\pm$ 0.08 & \citet{1966AJ.....71..709C} \\
 Stromgren $v$ & 6.775 $\pm$ 0.08 & \citet{1966AJ.....71..709C} \\
 Stromgren $b$ & 6.189 $\pm$ 0.08 & \citet{1966AJ.....71..709C} \\
 Stromgren $y$ & 5.80 $\pm$ 0.08 & \citet{1966AJ.....71..709C} \\
 Stromgren $u$ & 7.682 $\pm$ 0.05 & \citet{1994AandAS..106..257O} \\
 Stromgren $u$ & 7.695 $\pm$ 0.05 & \citet{1994AandAS..106..257O} \\
 Stromgren $u$ & 7.681 $\pm$ 0.05 & \citet{1994AandAS..106..257O} \\
 Stromgren $u$ & 7.672 $\pm$ 0.05 & \citet{1994AandAS..106..257O} \\
 Stromgren $b$ & 6.189 $\pm$ 0.05 & \citet{1994AandAS..106..257O} \\
 Stromgren $b$ & 6.204 $\pm$ 0.05 & \citet{1994AandAS..106..257O} \\
 Stromgren $b$ & 6.186 $\pm$ 0.05 & \citet{1994AandAS..106..257O} \\
 Stromgren $b$ & 6.183 $\pm$ 0.05 & \citet{1994AandAS..106..257O} \\
 Stromgren $v$ & 6.775 $\pm$ 0.05 & \citet{1994AandAS..106..257O} \\
 Stromgren $v$ & 6.781 $\pm$ 0.05 & \citet{1994AandAS..106..257O} \\
 Stromgren $v$ & 6.76 $\pm$ 0.05 & \citet{1994AandAS..106..257O} \\
 Stromgren $v$ & 6.756 $\pm$ 0.05 & \citet{1994AandAS..106..257O} \\
 Stromgren $y$ & 5.8 $\pm$ 0.05 & \citet{1994AandAS..106..257O} \\
 Stromgren $y$ & 5.82 $\pm$ 0.05 & \citet{1994AandAS..106..257O} \\
 Stromgren $y$ & 5.799 $\pm$ 0.05 & \citet{1994AandAS..106..257O} \\
 Stromgren $y$ & 5.799 $\pm$ 0.05 & \citet{1994AandAS..106..257O} \\
 Geneva $V$ & 5.794 $\pm$ 0.08 & \citet{1976AandAS...26..275R} \\
 Geneva $V1$ & 6.541 $\pm$ 0.08 & \citet{1976AandAS...26..275R} \\
 Geneva $B$ & 5.571 $\pm$ 0.08 & \citet{1976AandAS...26..275R} \\
 Geneva $B1$ & 6.651 $\pm$ 0.08 & \citet{1976AandAS...26..275R} \\
 Geneva $B2$ & 6.902 $\pm$ 0.08 & \citet{1976AandAS...26..275R} \\
 Geneva $U$ & 6.888 $\pm$ 0.08 & \citet{1976AandAS...26..275R} \\
 Geneva $G$ & 6.846 $\pm$ 0.08 & \citet{1976AandAS...26..275R} \\
 2Mass $Ks$ & 4.388 $\pm$ 0.0230 & \citet{2mass_book} \\
 DDO $48$ & 6.017 $\pm$ 0.05 & \citet{1981PDAO...15..439M} \\
 DDO $45$ & 7.02 $\pm$ 0.05 & \citet{1981PDAO...15..439M} \\
 DDO $42$ & 7.643 $\pm$ 0.05 & \citet{1981PDAO...15..439M} \\
 DDO $41$ & 7.658 $\pm$ 0.05 & \citet{1981PDAO...15..439M} \\
 DDO $38$ & 6.863 $\pm$ 0.05 & \citet{1981PDAO...15..439M} \\
 DDO $35$ & 7.767 $\pm$ 0.05 & \citet{1981PDAO...15..439M} \\
\enddata
\label{tab:sedmeas}
\end{deluxetable}

\end{document}